\documentclass[preprint,12pt]{aastex}
\newcommand{\kmax}{\ensuremath{k_{\mathrm{max}}}}
\newcommand{\dadra}{\texttt{DADRA}}
\newcommand{\wmap}{\textit{WMAP}}

\newcommand{\csq}{\ensuremath{\chi^2_\nu}}
\newcommand{\var}{\ensuremath{\mathrm{Var}}}
\shorttitle{\WMAP\ 5-year Beams}
\shortauthors{Hill et al.}
\slugcomment{Accepted for the Astrophysical Journal Supplement Series}
\begin{document}
\title{Five-Year Wilkinson Microwave Anisotropy Probe 
  (\wmap\altaffilmark{1})
  Observations:  Beam Maps and Window Functions}
\author{%
{R. S. Hill}    \altaffilmark{2},
{J. L. Weiland} \altaffilmark{2},
{N. Odegard} \altaffilmark{2}
{E. Wollack} \altaffilmark{3},
{G. Hinshaw} \altaffilmark{3},
{D. Larson}  \altaffilmark{4},
{C. L. Bennett} \altaffilmark{4},
{M. Halpern} \altaffilmark{5},
{L. Page}    \altaffilmark{6},
{J. Dunkley} \altaffilmark{6,7,8},
{B. Gold}    \altaffilmark{4},
{N. Jarosik} \altaffilmark{6},
{A. Kogut}   \altaffilmark{3},
{M. Limon}   \altaffilmark{9},
{M. R. Nolta}   \altaffilmark{10},
{D. N. Spergel} \altaffilmark{7,11},
{G. S. Tucker}  \altaffilmark{12},
and {E. L. Wright}  \altaffilmark{13}
}
\altaffiltext{1}{\wmap\ is the result of a partnership between Princeton
                 University and NASA's Goddard Space Flight Center. Scientific
                 guidance is provided by the \wmap\ Science Team.}
\altaffiltext{2}{{Adnet Systems, Inc.,  7515 Mission Dr., Suite A100, Lanham, Maryland 20706}}
\altaffiltext{3}{{Code 665, NASA/Goddard Space Flight Center,  Greenbelt, MD 20771}}
\altaffiltext{4}{{Dept. of Physics \& Astronomy,  The Johns Hopkins University, 3400 N. Charles St.,  Baltimore, MD  21218-2686}}
\altaffiltext{5}{{Dept. of Physics and Astronomy, University of  British Columbia, Vancouver, BC  Canada V6T 1Z1}}
\altaffiltext{6}{{Dept. of Physics, Jadwin Hall,  Princeton University, Princeton, NJ 08544-0708}}
\altaffiltext{7}{{Dept. of Astrophysical Sciences,  Peyton Hall, Princeton University, Princeton, NJ 08544-1001}}
\altaffiltext{8}{{Astrophysics, University of Oxford,  Keble Road, Oxford, OX1 3RH, UK}}
\altaffiltext{9}{{Columbia Astrophysics Laboratory,  550 W. 120th St., Mail Code 5247, New York, NY  10027-6902}}
\altaffiltext{10}{{Canadian Institute for Theoretical Astrophysics,  60 St. George St, University of Toronto,  Toronto, ON  Canada M5S 3H8}}
\altaffiltext{11}{{Princeton Center for Theoretical Physics,  Princeton University, Princeton, NJ 08544}}
\altaffiltext{12}{{Dept. of Physics, Brown University, 182 Hope St., Providence, RI 02912-1843}}
\altaffiltext{13}{{UCLA Physics \& Astronomy, PO Box 951547,  Los Angeles, CA 90095-1547}}
\email{{Robert.S.Hill@nasa.gov}}
\begin{abstract}
  Cosmology and other scientific results from the \wmap\ mission require
  an accurate knowledge of the beam patterns in flight.  While the
  degree of beam knowledge for the \wmap\ one-year and three-year
  results was unprecedented for a CMB experiment, we have significantly
  improved the beam determination as part of the five-year data release.
  Physical optics fits are done on both the A and the B sides for the
  first time.  The cutoff scale of the fitted distortions on the primary
  mirror is reduced by a factor of $\sim2$ from previous analyses.
  These changes enable an improvement in the hybridization of Jupiter
  data with beam models, which is optimized with respect to error in the
  main beam solid angle.  An increase in main-beam solid angle of $\sim
  1\%$ is found for the V2 and W1--W4 differencing assemblies.  Although
  the five-year results are statistically consistent with previous ones,
  the errors in the five-year beam transfer functions are reduced by a
  factor of $\sim 2$ as compared to the three-year analysis. 
  We present radiometry of the planet Jupiter as a test of the beam
  consistency and as a calibration standard; for an individual
  differencing assembly, errors in the measured disk temperature are
  $\sim 0.5\%$.  
\end{abstract}
\section{Introduction}
The \wmap\ mission has produced an unprecedented set of precise
and accurate cosmological data, resulting in a
consensus on the contents of the universe.  \wmap\ has
determined the age of the universe, the epochs of the key
transitions of the universe, and the geometry of the universe,
while providing the most stringent data yet on inflation.  At the
center of these advances is the angular power spectrum of the CMB,
which is the fundamental tool for studying the constituents and
density distribution of the early universe.

Characterizing the \wmap\ beams is crucial to interpretation of the
CMB power spectrum.  This paper, which describes the \wmap\ beam
analysis based on the five-year data set, is one of seven that
together describe the complete five-year \wmap\ analysis.  The results
from the suite of papers are summarized and set in context by
\citet{hinshaw/etal:prep}, which also describes the gain calibration,
data processing, and mapmaking.

\wmap\ observes in multiple microwave frequency bands, namely, K
($\sim 23$ GHz), Ka ($\sim 33$ GHz), Q ($\sim 41$ GHz), V ($\sim 61$
GHz), and W ($\sim 94$ GHz).  The sky is observed differentially via
two back-to-back offset Gregorian telescopes and associated
instrumentation, which are designated side A and side B.  The two
sides comprise ten independent sets of feed horns and radiometers,
called differencing assemblies (DAs): one each in the K and Ka bands,
two each in the Q and V bands, and four in W band
\citep{bennett/etal:2003}.  Each DA is designated by addition of a
digit to the name of the frequency band, giving the names K1, Ka1, Q1,
Q2, V1, V2, W1, W2, W3, and W4, respectively; thus, for example, Ka is
a frequency band, and Ka1 is the corresponding DA.

The terminology applied to beams can be subtle, and it reflects, to
some extent, the details of the particular analysis that is done.
Here, we give a brief overview.  For a fuller exposition of
beam-related concepts and notation, the reader is referred to
\citet{page/etal:2003e} for the main beams and
\citet{barnes/etal:2003} for the far sidelobes.  The \wmap\ optics are
described and analyzed in \citet{page/etal:2003}.

We can directly make a measured beam for any DA, either separately
for the A and B sides, or by averaging the two sides.  Similarly,
we can make beams for either of the two radiometers in a DA, which
measure orthogonal polarizations, or combine the two.  In a strict
sense, the word \emph{beam} means point-source response in
spherical coordinates, covering the full $4\pi$ steradians.  We
divide the full beam into two parts, which are measured
differently and treated differently in the analysis: the
\emph{main beam} and the \emph{far sidelobes}.  The main beam has
a radius of $3\fdg5-7\arcdeg$ depending on the DA.  The measured
main beam includes observations of Jupiter, while the measured far
sidelobes include in-flight observations of the Moon, as well as
pre-flight laboratory data.

The A- and B-side main beams can be predicted using physical models of
the \wmap\ optics.  Indeed, we go further and adjust the model
parameters in an iterative $\chi^2$ fit to arrive at \emph{model main
  beams}, or more simply, \emph{beam models}.  The instrument parameters
of interest are small surface distortions of the mirrors, especially
the two primaries.  Mirror distortions are modeled as
low-amplitude Fourier or Bessel modes added to the nominal mirror
shapes.  At intermediate angular scales within the main beams, the
models are actually more reliable than observations, so that we
combine models and observations to produce \emph{hybrid beams}.  

An additional variation in terminology is produced by our attempt to
reconcile the farthest outskirts of the model main beams with the
far-sidelobe observations of the Moon.  In this part of the analysis,
the parameter set of the best-fit model is augmented with extrapolated
small-scale distortions of the primary mirror to produce
\emph{augmented beam models}.

Scan strategy combines with the inherent geometry of each beam to
produce maps with effective beams that are nearly azimuthally
symmetric, when averaged over each year of observations.  The beam
analysis results in a symmetrized beam profile, which is
equivalent to a symmetrized point-spread function in optical
astronomy.  The transform of the beam profile in harmonic space is
termed a \emph{beam transfer function}, $b_\ell$.  A raw CMB power
spectrum is divided by $w_\ell=b_\ell^2$ to invert the filtering
done by finite-width beams, and the resulting beam-corrected power
spectra are used for fitting cosmological parameters; $w_\ell$ is
called the \emph{window function}.

Beam measurements consist of repeated scans over the planet Jupiter by
each DA, which occur as part of the standard \wmap\ full-sky observing
strategy, with no need for special observations
\citep{bennett/etal:2003}.  Jupiter is effectively a point source that
allows high-resolution sampling of each beam.  Because the sky is
covered completely in six months, every observing year includes two
Jupiter seasons, each lasting $\sim 50$ days. The data taken when
Jupiter is near the axis of each of the twenty main beams are
extracted from the time-ordered data (TOD) archive and analyzed
separately from the sky map processing.  Thus, the Jupiter data are
reduced in the same manner as the CMB data in terms of baseline
removal and gain calibration \citep{hinshaw/etal:prep}.  Each
brightness sample is labeled with the instantaneous position of
Jupiter's image in the A-side or B-side focal plane.  These data may
be either utilized in time-ordered form or accumulated into twenty
two-dimensional \emph{beam maps}.

The first-year beam analysis was described in \citet{page/etal:2003e}.
For each DA, the beams for the largely symmetric A and B optics were
measured independently.  Azimuthally symmetrized beam profiles were
fitted to the time-ordered data (TOD) of Jupiter using Hermite
functions as the basis.  The A- and B-side beams were averaged
to give one beam per DA.  Mean asymmetry corrections were produced by
a time integration of the beam orientation.

The first-year beam analysis also included the fitting of the detailed
shapes of the primary and secondary mirrors, motivated in part by the
fact that the cold, laboratory-measured rms of the primary mirror surface 
distortions did not
meet the pre-flight specification \citep[\S2.6]{page/etal:2003}.
Inputs to a physical optics modeling program, \dadra\
\citep{YRS:DADRA}, were varied iteratively to match Jupiter data.
This program requires four types of inputs:  (1) as-built coordinates
and Euler angles of the primary and secondary mirrors and the feed
horns, on both the A and the B sides; (2) as-designed feed horn
outputs expressed as a spherical-wave approximation; (3) as-built
primary and secondary mirror shapes; (4) perturbation coefficients for
small distortions of the mirror shapes, which were fitted as free
parameters.  In the first-year analysis, the fit was done only on the
A side and was used mainly to check the integral of the
peak-normalized beam, called \emph{beam solid angle}
\citep{page/etal:2003e}.

The three-year beam analysis was described in
\citet{jarosik/etal:2007}.  With the increase in the number of Jupiter
seasons, the physical optics fits became well enough constrained to
allow the model beams to be extrapolated below the noise level of the
data; thus, the models entered directly into the adopted beam
profiles.  Directly measured beam points based on Jupiter were
replaced with predicted values from a beam model if the corresponding
model point had a gain below some defined threshold
$B_\mathrm{thresh}$.  This merger, or hybridization, was done before
Hermite function fitting.  Again, the physical optics modeling was
applied only to the A side, and the result was transformed to be
applicable to the B side.  Also, residual departures from azimuthal
symmetry in the scan-averaged beams were computed for each DA and
found to be subdominant, except in Q band at $\ell \gtrsim 500$
\citep{hinshaw/etal:2007}.  Beam errors were computed to encompass the
overall uncertainty in the data, the modeling, and the transformation
from A side to B side.

The five-year beam analysis is the main subject of this paper.  An
additional four seasons of Jupiter observations have increased the
signal-to-noise ratio ($S/N$) of the beam data, allowing the
boundaries of the main beams to be expanded to lower gain levels.  The
scope of the physical optics fitting has been extended in two main
ways.  First, separate, complete fits are now done on both the A- and
the B-side optics.  Second, the cutoff scale of the fitted distortions
on the primary mirror is reduced by a factor of $\sim 2$, in effect
widening the beam area that is treated by the models.  The
hybridization step has been changed in that it is now optimized
statistically, by minimizing the error in the beam solid angle.  The
hybrid beam profiles are transformed directly to $b_\ell$, with no
intervening functional fit.  The improved $S/N$ and modeling result in
a $\sim 1\%$ increase in the solid angle of the adopted main beam for
several DAs, occuring mainly inside the old beam radius.  However, the
new beam transfer functions are consistent with those from the two
earlier analyses.

The five-year results include a revision to the \wmap\ full-sky
sidelobe sensitivity patterns.  Augmented beam models, which are
matched to early-mission observations of the Moon, are substituted for
a part of the sidelobe measurements taken under ambient ground-based
temperature and humidity.  In addition, the in-flight Moon
observations used directly in the sidelobe patterns are
recalibrated for bands Ka--W.  The resulting patterns are used in
\wmap\ sky map generation to correct for sidelobe pickup. 

An important test of the consistency of beam processing is radiometry
of the planet Jupiter.  These measurements, which should also be useful
in calibrating other microwave observations, are presented here,
together with radiometry of Mars and Saturn.

\begin{figure}
  \begin{center}
  \includegraphics[width=6.5in]{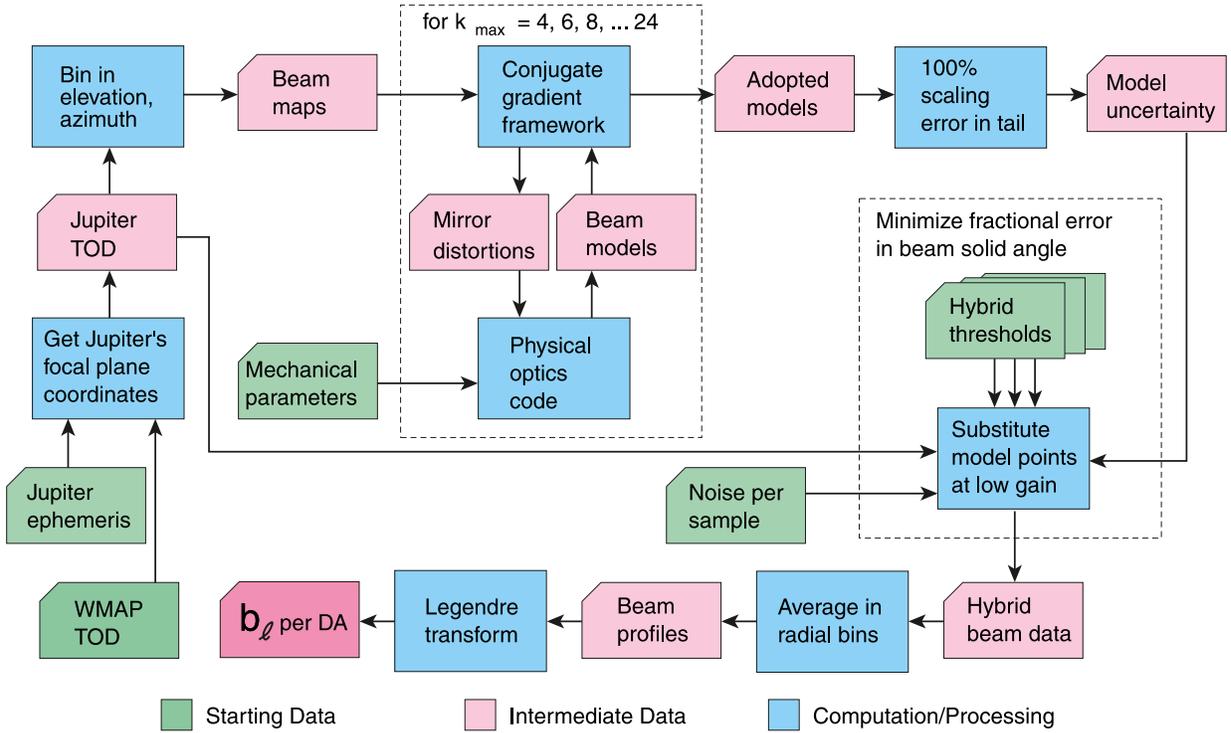}
  \caption{Flow chart of beam and window function processing.  Boxes
    are color coded as follows: \emph{green},  starting data;
    \emph{pink}, intermediate data; \emph{blue}, computations or 
    processing steps; \emph{magenta}, the result, $b_\ell$, which
    is the beam transfer function.  The window function, $w_\ell$,
    for power spectra $C_\ell$ involving a single DA, is $b_\ell^2$.  
    Dashed boxes enclose iterative algorithms.\label{fig:flow}}
  \end{center}
\end{figure}

A flowchart of the main beam processing, from TOD through window
functions, is shown in Figure \ref{fig:flow}.  This processing is
described in detail in \S\ref{sec:models}, which describes the fitting
of model beams, and in \S\ref{sec:beamwf}, which describes the
computation of beam transfer functions from Jupiter observations and
models.  Radiometry of selected planets is given in
\S\ref{sec:src}.  The conclusions of this study are summarized in
\S\ref{sec:conc}.
\section{Physical Optics Models\label{sec:models}}
\subsection{Beam Data}
The fundamental beam data are Jupiter measurements extracted from the
mission's TOD archive.  These measurements are differential like all
\wmap\ data, but the presence of Jupiter in only one of the two beams
of a given DA means that they are effectively single-dish, after
subtraction of a differential sky background.  Five-year \wmap\
full-sky maps, which omit Jupiter, are used to estimate this
background.  Also, measurements affected by a bright source other than
Jupiter in either set of optics are omitted.  The apparent measured
Jupiter temperatures are scaled to a standardized distance of 5.2 AU
and are binned by the position of Jupiter in a coordinate system
attached to the spacecraft.  Planetary coordinates are obtained from
the Jet Propulsion Laboratory ephemeris
DE200\footnote{\texttt{http://ssd.jpl.nasa.gov/}}\citep{standish:1990}.
\subsection{\dadra\ Modeling}
\subsubsection{Software Structure}
The physical optics modeling is the most computationally intensive
aspect of the \wmap\ beam analysis.  Individual fitting runs can take
days or weeks, even with the availability of numerous processors.  The
procedure is automated as much as possible to minimize user
intervention and false starts. 

Several pieces of software are used in concert.  Computation of a beam
model from a specification of the mirrors and of a single feed is done by
a Fortran program called \dadra\ \citep{YRS:DADRA}.  The overall
framework of the fitting algorithm is embodied in an Interactive Data
Language (IDL) code\footnote{IDL is licensed by ITT Visual
  Information Solutions, Boulder, Colorado.}, which performs a
conjugate-gradient $\chi^2$ minimization driven by the residuals
between beam models and the Jupiter data.  Requests for beam
computations are made by the IDL program, and they are passed to
\dadra\ by intermediary shell scripts running continuously in the
background.  This architecture enables the computation to be spread
over multiple processors, with each processor running an instance of
\dadra. 

The minimization software is in most respects the same as that used in
the first two \wmap\ beam analyses \citep{page/etal:2003e,
  jarosik/etal:2007}.   In the one- and three-year analyses, ten
processors of one computer were used concurrently to compute the beams
for a given mirror distortion.  However, for the five-year fits,
upgrades to the IDL and shell code enable the use of multiple
computers.  Typically, the fit uses four to six clustered Silicon
Graphics Origin 300 machines, each with 32 processors and 32 gigabytes
of memory in a non-uniform memory access (NUMA) architecture.  With
these resources, 12 to 18 sets of ten beam models each can be computed
simultaneously.
\subsubsection{Parameters of Fit}
The coordinates and Euler angles of the \wmap\
optical components are furnished to \dadra\ as constants, as are the
measured mirror shapes and the beam patterns of the feed horns.  These
constants are results of the pre-flight structural thermal optical
(STOP) performance analysis \citep{page/etal:2003}.  The
fit is done by varying a set of mirror distortion parameters, which
are defined as small-amplitude Fourier or Bessel modes added to the
input shape.  The emphasis in the fitting is on the primary mirror,
because in pre-flight laboratory measurements, the shape of the
secondary was less susceptible to temperature changes.
Also, the illumination pattern occupies a
much larger proportion of the primary mirror than of the secondary
\citep[Figure 2]{page/etal:2003}.  For these reasons, the perturbation space
chosen to characterize the primary mirror encompasses more modes than
that for the secondary.   Moreover, Fourier modes in $x$ and $y$ are a natural
choice because the morphology of the distortions is
dominated by a rectangular frame that is part of the backing structure.
On the other hand, the mount of the secondary is cylindrically
symmetric with a more extensive backing structure, consistent with
fitting to a restricted set of Bessel modes.

The primary distortions have the form
\begin{eqnarray*}
  \delta Z & = & \sum C(k_x,k_y) \cos\left({2 \pi k_x x\over L}
  + {2 \pi k_y y \over L} \right)\ \\
  & & + \sum S(k_x,k_y) \sin\left({2 \pi k_x x\over L}
  + {2 \pi k_y y \over L} \right)\ \\
  & & + \sum_{\mathrm{special}} A_{\mathrm{sp}} F_{\mathrm{sp}}(x,y).
  \label{eq:ppar}
\end{eqnarray*}
The distortion modes are characterized by the spatial frequency
indexes $k_x$ and $k_y$, where $k=1$ corresponds to $L=280$ cm, which
is twice the width of the mirror.  The reason for setting $L = 280$~cm
rather than $140$~cm is to remove the requirement that the solution be
periodic on the circumscribed square; in particular, an approximate
tilt of the $140$~cm mirror is expressed naturally as half a sine wave
with a $280$~cm period.  Some of the Fourier modes specified in this way
are redundant, so the set that is used in the fit is culled to contain
only even values of $k_x$ and $k_y$, with $k_x>0$.  A few additional
modes with $k_x$ and $k_y$ of $\pm 1$ are the ones that
represent tilt.  Special
distortion modes, $F_{\mathrm{sp}}(x,y)$, are also allowed, with
amplitudes $A_{\mathrm{sp}}$.   One special
mode is a scalar offset of the whole mirror surface.
A second special mode is simply a
map of the mirror surface as measured pre-flight;
in practice, this mode plays no significant role in the fit.
The phase and
strength of the modes are specified by the sine and cosine amplitudes,
$S(k_x,k_y)$ and $C(k_x,k_y)$, and the amplitudes of the special
modes, $A_{\mathrm{sp}}$.  Below, we make use of the power per
mode, $P(f) = C(k_x,k_y)^2 +
S(k_x,k_y)^2$, where $f$ is the spatial frequency in
$\textrm{cm}^{-1}$ and $f^2=(k_x/280)^2 + (k_y/280)^2$.

The three-year fit, carried out for the A side only, reached a maximum
spatial frequency index of $\kmax=12$, corresponding to a distortion
wavelength of $280/12 \sim 23$ cm.  By contrast, the five-year fit reaches
$\kmax=24$, or $\sim 12$ cm, for both the A and the B sides.  The
number of Fourier modes goes as $\kmax^2$, so the five-year
fits include $\sim 400$ modes on each primary as compared to $\sim
100$ modes in previous analyses.  This extension in $k$ space means
that the primary mirror distortions are fitted nearly to the
surface correlation length measured in the laboratory under cold
conditions, i.e., 9.3 cm on the A side and 11.4 cm on the B side
\citep{page/etal:2003}.

The secondary mirror distortions are described by Bessel functions:
\begin{equation}
  \delta Z = \sum_{n,k}  J_n(\nu_{n,k}\rho/L)[ C_{n,k} \cos(n \phi)
    + S_{n,k} \sin(n \phi) ]\ ,
  \label{eq:spar}
\end{equation}
where $\rho$ and $\phi$ are cylindrical coordinates and $L$ is
the radius of the mirror.  The $k$th alternating zero of $J_n,
J^\prime_n$ is denoted $\nu_{n,k}$; zeroes of $J_n$ have $k$ odd,
and zeroes of $J^\prime_n$ have $k$ even.  Inclusion of the zeroes of
$J^\prime_n$ removes any constraint on the edge of the mirror.  The
fitted parameters are $C_{n,k}$ for $n \geq 0$ and $S_{n,k}$ for
$n \geq 1$.  Two different pairs of $(n_{\mathrm{max}},
\kmax)$ are used:  $(1, 3)$, resulting in nine parameters, or $(2,
6)$, resulting in 30 parameters.
\subsubsection{Fitting Method\label{sec:method}}
The optimization is done by a modified conjugate gradient method,
a deterministic descent into a $\chi^2$ valley.  Avoidance
of local $\chi^2$ minima is attempted by exploiting the Fourier
description of the primary mirror distortions.  The largest scale
distortions are fitted first, and each result is used as a starting
point for the next fit, in which finer scale modes are included.
At each stage, distortions at the large scales that have already
been fitted are not held constant, but rather, they are refitted
together with the small-scale distortions that are newly included.

As defined above, the primary mirror modes do not compose an orthogonal
basis.  However, orthogonality is desirable in order to make the fit as
efficient as possible.  Consequently, the primary mirror modes are not
fitted directly, but first are orthogonalized with respect to the area
inside the circular boundary of the mirror, using a modified
Gram-Schmidt method.  When the $\kmax$ of the included Fourier modes is
increased, a new orthogonalization is performed that generates a
completely new set of linear combinations of the Fourier and special
modes.  The conjugate gradient algorithm therefore navigates in a space
consisting of two groups of parameters:  (1) the amplitudes of the
orthogonalized primary mirror modes, and (2) the amplitudes of the
Bessel modes on the secondary.

Each time $\chi^2$ is calculated, two types of adjustment are made to
the model beams.  First, the pointing in the coordinate frame attached
to the spacecraft is matched to that implied by the Jupiter
observations.  Second, the peak sensitivity of the model is scaled to
match the peak observed Jupiter temperature.  The pointing adjustment
may be done for the ten beams as a group, without altering their mutual
displacements, or it may be done for each beam separately.  However,
the peak scaling is always done separately for each beam.

These adjustments are not parameters of the fit, because corresponding
dimensions of $\chi^2$ space do not exist.  Rather, their purpose is
to absorb errors in the input coordinates and angles of optical
components and prevent them from being projected into the mirror
distortions.  Ideally, this problem would be avoided by solving a more
complicated problem, i.e., by directly fitting the mechanical
parameters of the \wmap\ components.  However, a full set of
mechanical parameters would include many degeneracies with respect to
the beam morphology.  By limiting the
parameter set to $\sim 100-400$ mirror perturbation modes and ignoring
``nuisance'' information, we converge on acceptable values of
$\chi^2$.

For a given fitting run, either the primary or secondary mirror
parameters can be held constant at the starting values.  The DA
microwave frequencies can also be fitted as parameters, but are
held constant in practice, since they are accurately determined.

The fits for side A begin with the inherited three-year solution
\citep{jarosik/etal:2007}.  The final run for each value of $\kmax$ is
listed in Table~\ref{tab:hist-a}.  An indication of the quality of fit
after each step is given by the $\csq$ column in the table.

The B side of the instrument is characterized by an overall shift of
the ten beam pointings in relation to their pre-flight positions, by
$\sim 0\fdg1$.  This shift complicates the fitting strategy.  The
overall fitting history for side B comprises several different
sequences of fits.  The most important sequence, leading to the adopted
beams, is similar to the A-side fitting sequence, in that $\kmax$ is
increased in stages, with the secondary mirror distortions held
constant after being fitted early in the sequence (Table
\ref{tab:hist-b}).  In other sequences, a different form of the
secondary was tried, the floating shift in elevation and azimuth was
disabled, or the secondary alone was fitted from various initial
conditions.  None of these variations improved $\chi^2$ for the
resulting beams, as compared to the adopted fitting sequence.

\begin{figure}
  \begin{center}
    \includegraphics[height=7in]{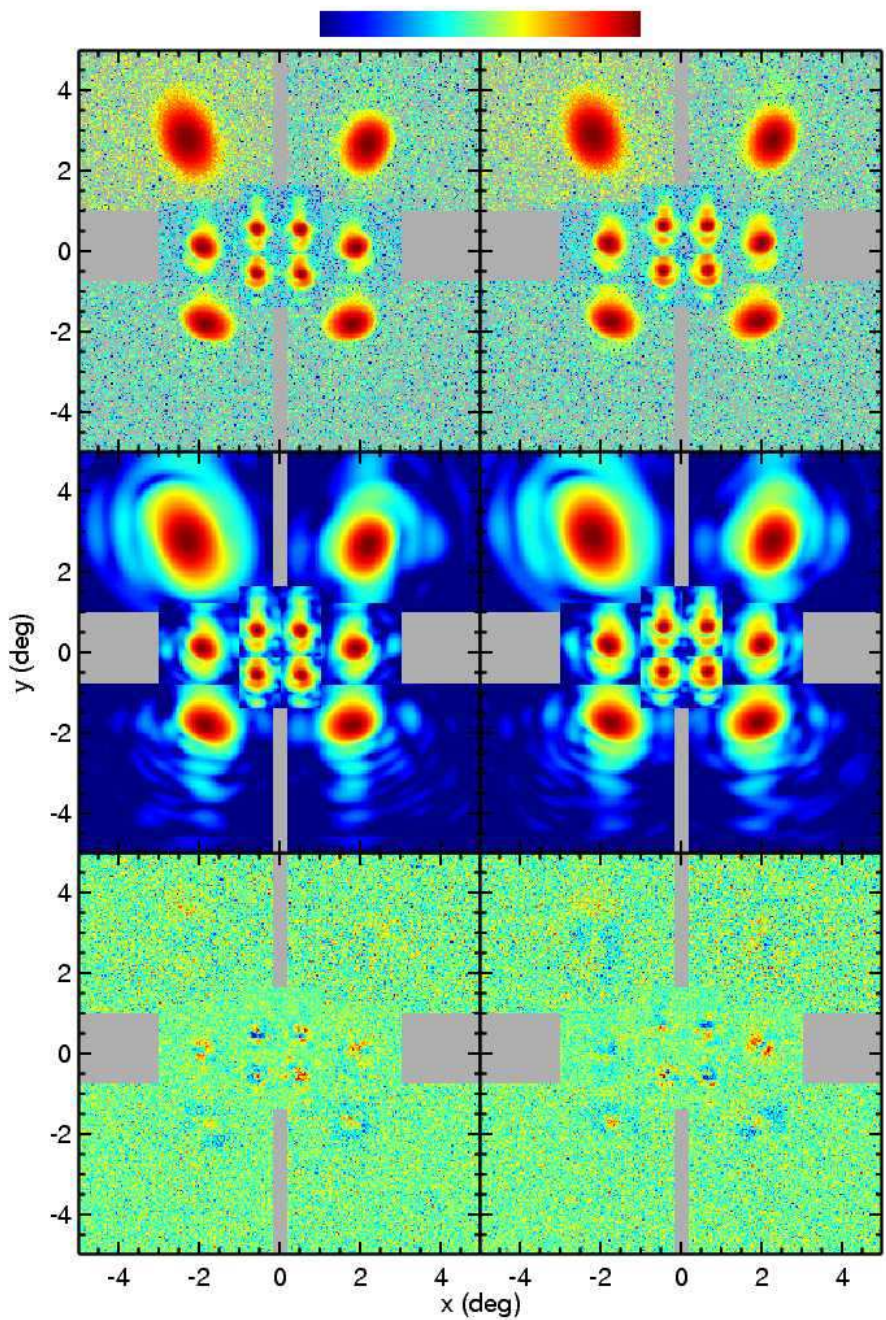}
    \caption{Beams in the \wmap\ focal plane for side A (left) and
      side B (right).  The top panels show the measured beams, the middle
      panels show the beam models, and the bottom panels show the
      residuals.  In the top four panels, each beam is scaled to its
      maximum (\emph{red}) and plotted logarithmically to a level
      of $-40$ dB (\emph{blue}).  For the bottom panels, each beam's
      residual is shown linearly as 100(data$-$model)/beam peak.
      The scales are $\pm 10\%$ for K1; $\pm 5\%$ for Ka1, Q1, and Q2;
      $\pm 3\%$ for V1 and V2; and $\pm 2.5\%$ for W1--W4.  A similar
      depiction of the A side only for three years of data is in
      Figure 9 of \citet{jarosik/etal:2007}.
      \label{fig:abresid}}
  \end{center}
\end{figure}

The fitted beams and residuals for sides A and B are
shown in Figure \ref{fig:abresid}, which can be compared
to Figure 9 of \citet{jarosik/etal:2007}.
\subsubsection{Instrument Parameter Results}
\begin{figure}
  \includegraphics[width=6in]{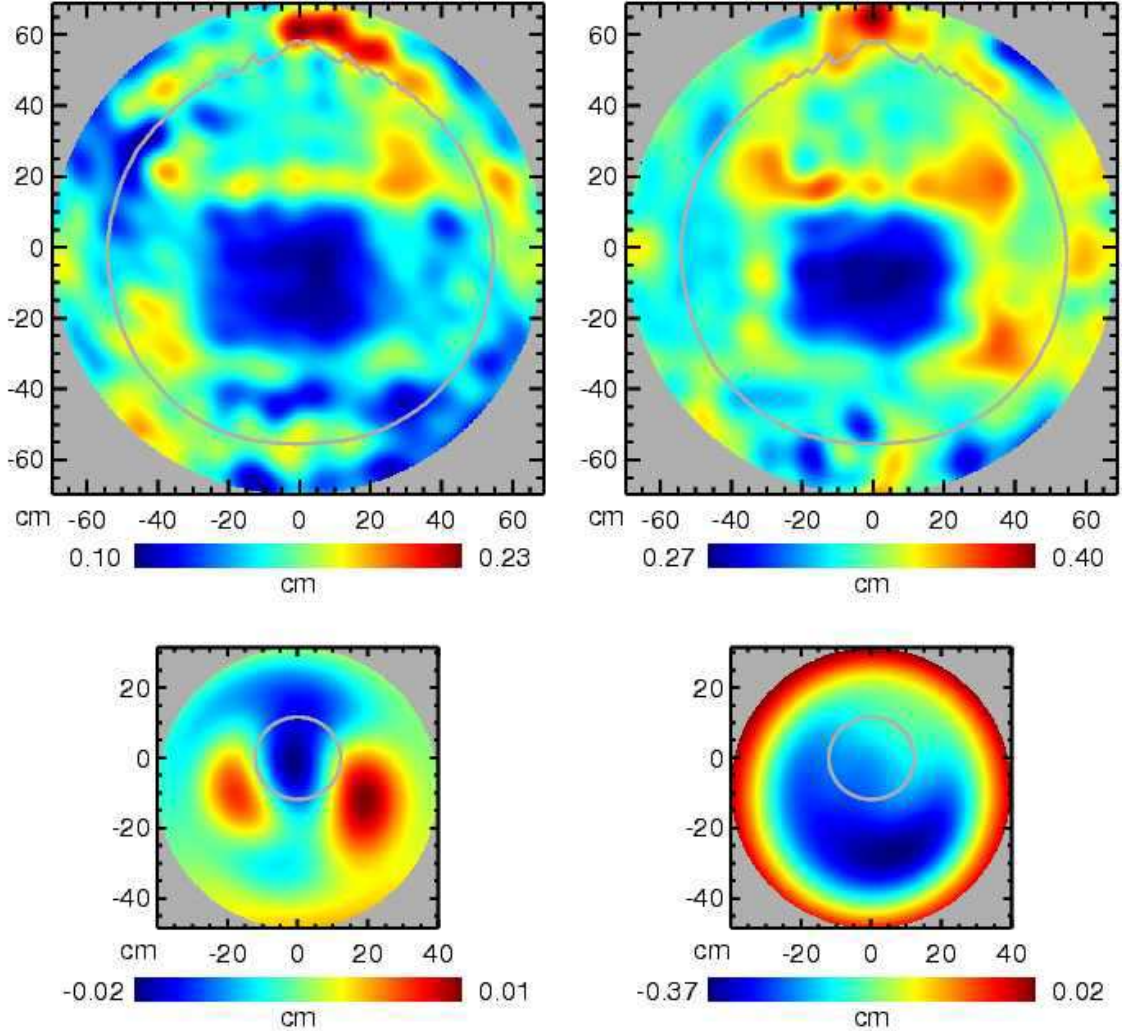}
  \caption{Fitted distortions of the A-side (left) and B-side (right)
    mirrors with respect to nominal shapes.  $Y$ axis is negative
    in the sky direction and positive toward the main spacecraft structure. 
    Top row:
    primary mirrors.   The dominant feature of the primary mirror distortions is the
    central rectangle, corresponding to a frame that is part of the
    backing structure.  Hints of the stiffening lugs in the backing
    structure may also be seen around the edges.   Bottom row: secondary mirrors.  The mirrors are
    constrained only where they are substantially illuminated by the
    feed horns \citep{page/etal:2003}.  Thus, for example, the secondary
    mirror for the B side appears as a bull's-eye partly because the fit
    is only constrained near the center.  \emph{Gray line}:  contour
    of the mean W band illumination function $-15$ dB from the peak.
    Although the mirrors are
    elliptical in outline, these plots are circular.  The reason is that
    the distortions are parametrized as displacements along the axis of
    a circular cylinder containing the mirror boundary.    \label{fig:abmirr}}
\end{figure}

False-color renditions of the final A and B side mirror surface fits
are shown in Figure \ref{fig:abmirr}.  Two natural length scales for
the surface of the primary mirrors are $0.5$ cm for a hexagonal mesh
that composes one layer of each mirror, and $30$ $\mu$m for the
correlation length of the reflector surface roughening, which was done
to diffuse visible solar radiation \citep{page/etal:2003}.  Both of
these length scales are too small to be probed either by the direct
fitting of the main beam or by the sidelobe observations of the Moon.
The main feature of each fitted primary mirror figure is the backing
structure, dominated by members that form a rectangular frame near the
center of the mirror.  In the center part of this rectangle, the
primary mirror appears to be depressed by $\sim 0.5-1$ mm.  Also seen
are hints of the stiffening lugs near the edge of each backing
structure.  The rms distortions of each primary mirror model are
$\sigma_z = 0.023$ cm and $0.022$ cm for the A and B sides,
respectively.  Pre-flight cold-measured values on the real mirrors, as
extrapolated to the flight temperature of 70 K, were $\sigma_z = 0.023$
cm and $0.024$ cm, respectively \citep{page/etal:2003}.

For the A side, the measured centroids of Jupiter beam data are
displaced by $\sim 0\fdg03 \pm 0\fdg02$ from the nominal pre-flight
beam positions on the sky, where the error term is the $1\sigma$
scatter among beams.  However, for the B side, the corresponding
displacement is $0\fdg13 \pm 0\fdg03$.  The $\chi^2$ computation in
the beam-fitting algorithm includes a floating elevation-azimuth
adjustment that is intended to soak up such discrepancies without
converting them into parameters of the fit.  For the final adopted
beam models, the floating displacement amounts to $0\fdg09$ in
combined elevation and azimuth for side A, and $0\fdg21$ for side B.
The difference between these two values agrees with the raw pointing
difference between the beams on each of the two sides.

We emphasize that these displacements are unrelated to the estimated
pointing errors of $<10\arcsec$ in the \wmap\ TOD
\citep{jarosik/etal:2007}.  The A-side and B-side boresight vectors
are accurately determined from flight data as part of the TOD
processing and are not influenced either by the beam fitting or by
pre-flight predictions.  The \wmap\ pointing model is described in
\citet{limon/etal:prep}. 

The mirrors are constrained only where they are substantially
illuminated by the feed horns; see \citet{page/etal:2003}, Figure 2. 
Thus, for example, the secondary mirror for the B side appears as a
bull's-eye partly because the fit is only constrained in the center
(Figure \ref{fig:abmirr}).  In actuality, the fitted shape consists
mostly of a tilt of $\sim 0\fdg25$; however, apparent mirror tilts
reflected in the fitted parameters are difficult to interpret because
they are coupled to the floating elevation-azimuth offsets.

The polarization characteristics of the main beam models are
consistent with previous results.  The morphology of the co- and
cross-polar components of both the A- and the B-side models is similar
to that found for the A side in the three-year analysis.  Similarly,
cross-polar suppression, as calculated from peak model values, is
within $\sim 0-2$ dB of previously reported A-side values, depending
on the DA.  However, for Q, V, and W bands, the polarization isolation
of the orthomode transducer (OMT) dominates the end-to-end cross-polar
response, which was measured pre-flight \citep{jarosik/etal:2007}.
\subsection{Extrapolation to Small Distortion Scales \label{sec:moonextrap}}
The modeling of primary mirror distortions with $k$ as high as 24,
which affect the beam at relatively wide angles, raises the
possibility of comparing \dadra-computed main beams to the innermost
parts of the far sidelobe patterns, which are obtained from
observations of the Moon \citep{barnes/etal:2003}.  

The beams at angles greater than $5\arcdeg - 10\arcdeg$ from each boresight
may be affected by unmodeled primary mirror distortions with $24 <k
\lesssim 250$.  However, extending the models to fit these distortions
directly is computationally unmanageable, because the required number
of Fourier modes is of order $10^4$, which is $\sim 100$ times
the number of modes in our normal fits.  Nevertheless, the observed
sidelobe data provide a constraint on the contribution of such modes
to the primary mirror surface shapes, and hence to the main beams.  To
apply this constraint, we need an appropriate choice of sidelobe data
together with a method for extending each main beam model to the
low gain levels outside the main beam radius.

The choice of sidelobe data is important, because the full-sky
sidelobe patterns are dominated by features that are not captured in
the beam models.  The \wmap\ sidelobe patterns are depicted in Figure
2 of \citet{barnes/etal:2003}.  The most vivid features are formed by
reflections from the radiator panels, by diffraction around the edges
of the primary mirrors, and by reflection from the focal plane
assembly.  Despite their striking appearance in the referenced figure,
these features are at least $40-60$ dB below the peak gain of each
main beam; however, they must still be excluded from the comparison.

Only a small region of each sidelobe map, a smooth region in the
``shadow'' of the primary mirror, is suitable for comparison with the
main beams.  Although we choose the boundary of this region as
conservatively as possible, the choice is subjective and a source of
systematic error.  Essentially, we draw the boundary according to a
combination of sidelobe morphology and angular proximity to the main
beam.  However, the morphological criterion by nature cannot exclude
an extraneous component of sidelobe response that happens to be
smooth, such as might arise from a diffuse reflection off the top
of the structure that holds the feeds.  For this
reason, the sidelobe patches chosen for the comparison are regarded as
upper limits. 

Additionally, these radiometric observations cannot be calibrated as
well as the CMB data, primarily because they were taken when the
spacecraft was thermally unsettled, during the phasing loops between
the Earth and the Moon.  The instrumental gain in this part of the
mission is estimated to be known to $\lesssim 10\%$.  A total of $4.3$
days of Moon data were obtained covering $\sim 1.2\pi$ sr of sidelobe
area.  The calibration standard for the Moon observations is the COBE
DMR model of lunar microwave emission as a function of phase angle
\citep{bennett/etal:1992a}.

\begin{figure}
  \includegraphics[width=7in]{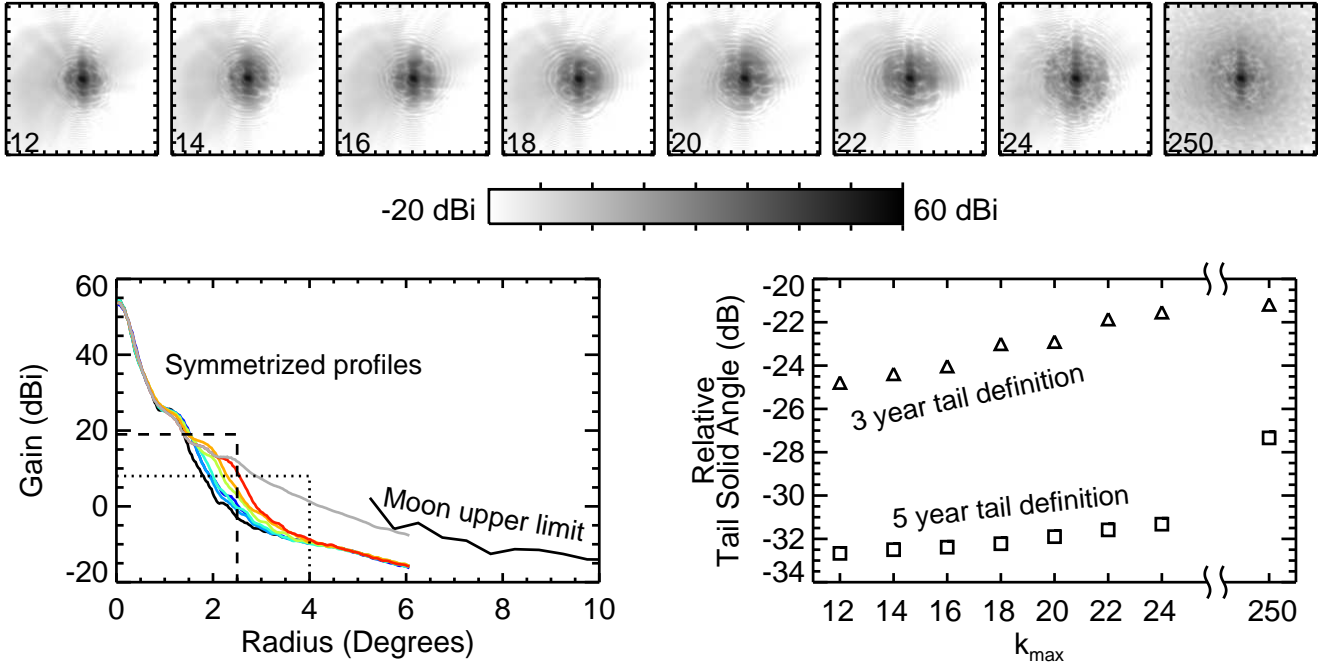}
  \caption{Growth of solid angle in model beams as a function of $\kmax$, 
    for the V2 DA on the A side.  \textbf{Top}:  logarithmically
    scaled images of the model beam pattern as fitted for $\kmax= 12$,
    14, 16, 18, 20, 22, and 24, respectively, together with a beam
    that combines the $\kmax=24$ fit with random-phase modes
    extrapolated to $k=250$.  Axis tick marks are at $1\arcdeg$
    intervals.  \textbf{Bottom left}: azimuthally averaged beam
    profiles for the models pictured, in units of gain relative to
    isotropic.  \emph{Indigo--red}: profiles of the seven beam models
    from $\kmax=12$ through $\kmax=24$, respectively.  \emph{Gray}:
    $\kmax=250$ extrapolation result.  \emph{Black}:  Upper limit on
    the main beam sensitivity from Moon observations, obtained for
    side A by integrating over positive pixels in the differential sidelobe
    response pattern.  The five-year model tail, which is a feature of
    the two-dimensional beam pattern, is the part of the beam that is
    both inside the transition radius and below the hybridization
    threshold (\emph{dotted line}; see \S\ref{sec:beamprof}).  The
    hybridization threshold and transition radius from the three-year
    analysis are indicated by the \emph{dashed line}.  \textbf{Bottom
    right}:  model tail solid angle as a function of $\kmax$, relative
    to the total solid angle inside the transition radius; squares,
    five-year; triangles, modified tail of five-year models, using
    three-year threshold and transition radius.  Fitting to $\kmax=24$
    rather than $\kmax=12$ increases by a factor of $\sim 2$ the solid
    angle of the model tail as defined by three-year main beam limits.
    But note that the difference between the various fits is $\sim
    0.1\%$ of the total beam solid angle for the 5 year data.
    \label{fig:kmx}}
\end{figure}

In order to achieve a comparison with the Moon observations, the
spatial frequency of the modeled primary mirror distortions is
pushed to as high a value as possible.  Figure \ref{fig:kmx} (lower
left panel) shows radial profiles of the symmetrized main beam models
in comparison to the sidelobe sensitivity pattern for one example
beam, V2 on the A side.  The sensitivity profile at intermediate
angles of $1\fdg5-2\fdg5$ is directly related to $\kmax$ of the fit,
as seen also in the grayscale images of the model beams (top row).  A
natural hypothesis is that an even greater increase in $\kmax$ might
give a model that joins smoothly to the Moon data.  Unfortunately, the
direct fitting algorithm cannot accommodate an indefinite increase in
$\kmax$, because the number of Fourier modes goes as $\kmax^2$.

\begin{figure}
  \begin{center}
    \includegraphics[height=6in]{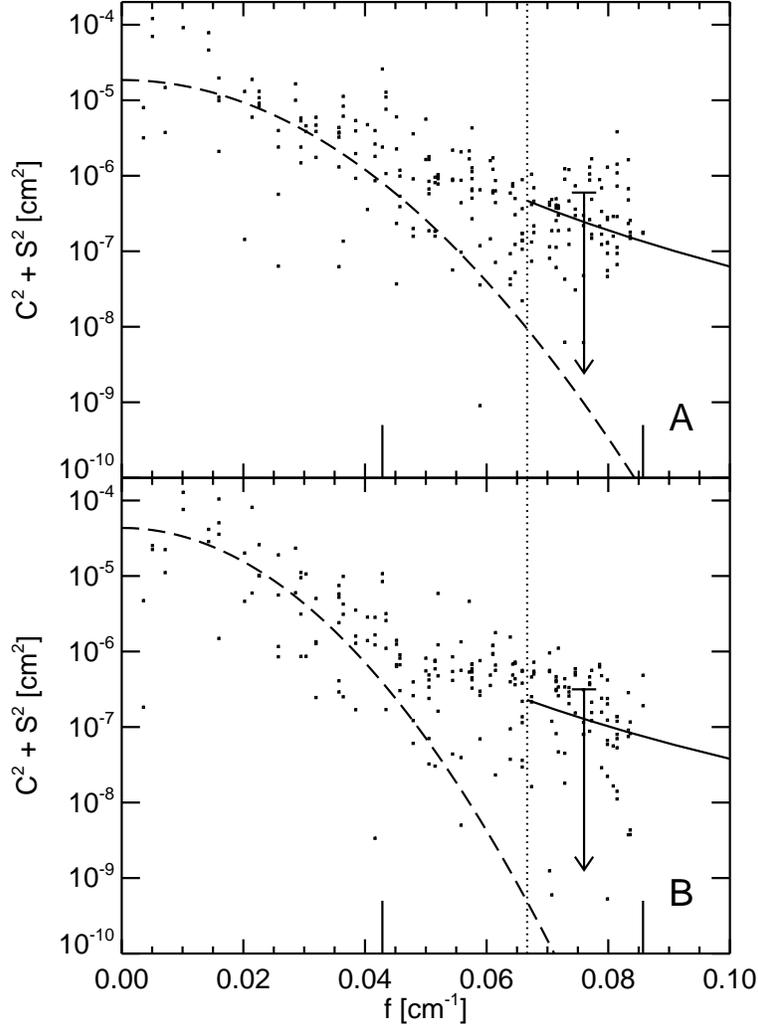}
    \caption{Power spectra of A-side (top) and B-side (bottom) primary
      mirror distortions as a function of spatial frequency on the primary
      mirror surface, $f=k/(280\ \mathrm{cm})$, where $k$ is the spatial
      frequency index used in the physical optics fits.  Vertical bars
      on the $f$ axis indicate $f$ corresponding to $k=12$ and $k=24$.  
      Solid lines: extrapolated
      power-law distortion spectra with slopes fitted by comparison to
      Moon sidelobe data, namely, $\alpha=4.95$ for the A side and 
      $\alpha=4.43$ for the B
      side.  In practice, these extrapolated distortions are used to
      update the sidelobe response patterns, but not to model the main
      beams.  The error bars and upper limits show the mean 
      absolute deviation about the mean of
      points with length scales less than $15$ cm, indicated by the dotted
      vertical line. Dashed curves:  power spectra of primary mirror
      distortions from ground-based laboratory measurements of the
      surface, assuming a Gaussian form for the two-point correlation
      function, with correlation length 9.3 cm for the A side and 11.3 cm
      for the B side; normalized to  points with $f<0.05$ cm$^{-1}$ (top)
      and $f<0.04$ cm$^{-1}$ (bottom). \label{fig:ps-ab}}
  \end{center}
\end{figure}

To cope with this difficulty, an approximation is used for Fourier
modes with $25 \leq k \leq 250$.  Power spectra, $P$, of the fitted
A-side and B-side primary mirror distortions are shown in Figure
\ref{fig:ps-ab} as a function of spatial frequency, $f=k/280$
cm$^{-1}$.  The form of the power spectrum expected from ground-based
measurements of the mirrors is also shown, under the assumption of a
Gaussian two-point correlation function \citep{page/etal:2003}.  As
the spatial frequency increases, $P$ decreases.  Our approach is to
extrapolate the power spectrum $P(k)$ to smaller scales assuming a
power-law form, $P \propto (1/k)^\alpha$, with $3 \lesssim \alpha
\lesssim 6$.  Random phases are used to convert the extrapolated
spectrum to sine and cosine amplitudes.

A grid of beam models is assembled as a function of two variables: the
slope $\alpha$, and the random number seed $s$ that selects the phases
used for the extrapolated Fourier modes.  Separately for each $s$, a
$\chi^2$ minimization is used to fit the beam models to the Moon
sidelobe data as a function of the slope $\alpha$.  The sidelobe maps
are in HEALPix format \citep{gorski/etal:2005} with
$\texttt{nside}=512$, and the beam models are resampled onto the
HEALPix grid.  The value of $\chi^2$ is computed from measured and
predicted gains, $g$, as $\sum(g_{\mathrm{pred}}-g_{\mathrm{Moon}})^2/
\sigma_{\mathrm{Moon}}^2$, where the sum is over pixels in the region
of overlap between the beam model and the Moon data.  The fit uses all
of the Q, V, and W DAs together.  For each side, the fitted values of
$\alpha$ from five values of the seed $s$ are averaged to get the
adopted slope.  Using this slope value, five new beam models are
computed with the original phases, and these models are averaged.  The
result is termed the \emph{augmented beam model}.

\begin{figure}
  \begin{center}
    \includegraphics[height=7in]{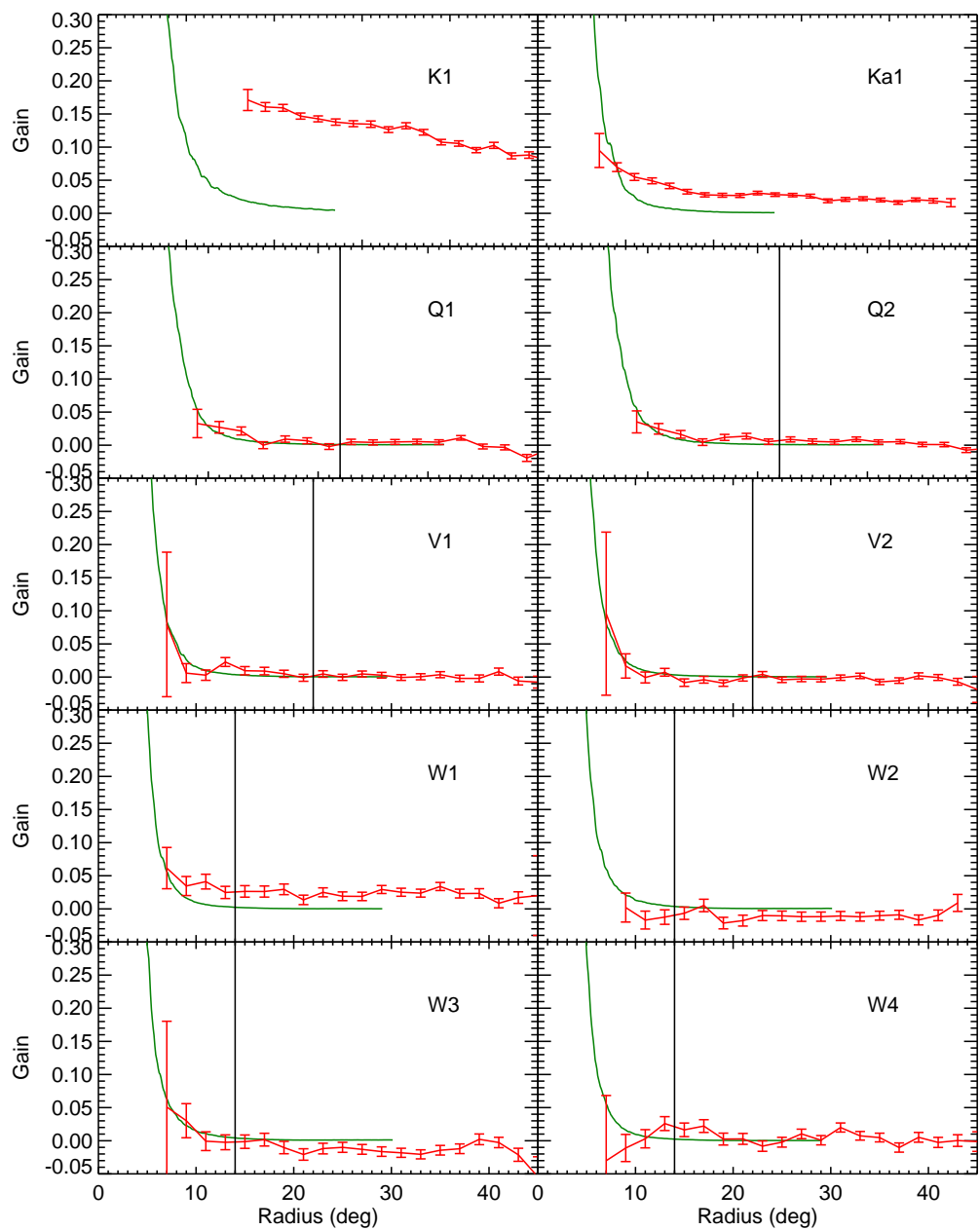}
    \caption{A-side augmented beam profiles (\emph{green})
      compared to Moon sidelobe data (\emph{red}).  K1 and Ka1 appear to
      be dominated by diffuse reflection rather than the extended main
      beam, and so are excluded from the fit.  Conspicuous DA-to-DA
      differences are seen in the quality of the fit, e.g., W1 and W2
      as compared to V1 and V2.  Contamination of the fit by diffuse
      reflected light cannot be ruled out even in DAs other than K1 and
      Ka1; thus, the Moon data are best considered as upper limits.
      Vertical line: maximum radius of Moon data included in fit, for
      DAs Q1--W4.
      \label{fig:moon-a}}
  \end{center}
\end{figure}

Figure \ref{fig:moon-a} shows radial profiles of the A-side augmented
beam models compared to the fitted subset of Moon data.  The K1 and Ka1
beams are not used in the fit because of the relatively strong diffuse
light that is seen as the bright profile in the top two panels of
Figures 6 and 7.  This component, which may result from reflection off
the focal plane assembly, is seen in Figure  2 of
\cite{barnes/etal:2003}, where it appears as a haze in the region
above the main beam.  In addition, the Q, V, and W bands appear
subject to systematic errors depending on the individual DA.

For CMB analysis, the main implication of the augmented models is an
increase in main beam solid angle as compared to the ordinary fitted
models with $\kmax =24$.  The effect on the ``tail'' part of the main
beam model is illustrated in the bottom right panel of Figure
\ref{fig:kmx}.  To compute accurately the effect that the augmented
models would have on the symmetrized beam profiles used to compute
$b_\ell$, a hybridization with Jupiter data is required, as described
below (\S\ref{sec:hybrid}).  If augmented rather than ordinary
$\kmax=24$ beam models are used in the hybridization, the resulting
increase in main beam solid angle is just $\sim0.1-0.3\%$
depending on DA.

There are two arguments for treating this main-beam solid angle
increase as an upper limit. One argument is the one already made
above, namely, that the Moon data in all bands may include a diffuse
reflected component in addition to the extended main beam response, as
is apparent for K1 and Ka1, and which our procedure cannot exclude.
The other argument invokes the thermal nature of the CMB
power spectrum, which requires that the power spectrum be the same in
all microwave frequency bands.  Section \ref{sec:btf} below shows that
the ordinary models with $\kmax=24$ maintain a tighter consistency of
the CMB $C_\ell$ across Q, V, and W than do the augmented models.
Consequently, the adopted 5-year beam transfer functions incorporate
only the ordinary models, whereas the augmented models are used to
characterize the innermost part of the sidelobe response
(\S\ref{sec:sl}).
\section{Beams and Window Functions\label{sec:beamwf}}
\subsection{Hybridization \label{sec:hybrid}}
To mitigate sensitivity of the window functions to observational
noise, we use a beam hybridization technique similar to that employed
in the three-year data analysis \citep{jarosik/etal:2007}.  In this
method, a hybrid beam is constructed for each DA on each side by
combining Jupiter observations with the physical optics models.
Jupiter observations are used in the central portions of the beam
where $S/N$ is high.  Model points are substituted
for the data in the outlying regions of low signal, called the tail.

The five-year analysis differs from that of \cite{jarosik/etal:2007} in
the method for choosing the hybridization threshold,
$B_\mathrm{thresh}$, which defines the tail region.  In the three-year
analysis, the threshold for each DA was chosen to replace
noise-dominated parts of the beam with model values, in order to
facilitate fitting the beam profile with a smooth function.  However,
the improvements in the five-year data and modeling open the
possibility of extrapolating the main beam to wider angles and
subsuming more of the full-sky beam solid angle into the main-beam
treatment, rather than the sidelobe pattern.  As a result, lower-signal
parts of the beam are included in the beam transfer functions, and we
require an explicit optimization of $S/N$ in the hybrid
beams.

The effect of the beam tail on science occurs through the normalization
of the CMB power spectrum, $C_\ell$.  An increase in main beam solid
angle raises the beam-corrected $C_\ell$ by a constant factor for $\ell
\gtrsim 100$.  Hence, the error in the solid angle is a convenient
indicator of the error induced in the high-$\ell$ part of the CMB power
spectrum via the hybridization.  The solid angle error is therefore a
natural figure of merit for optimizing $B_\mathrm{thresh}$.

A grid of simulations is run that evaluates solid angle error in the
hybrid beam as a function of threshold level.  Let the true
solid angle of a given beam be $\Omega$, and the solid angle of
the hybrid beam be $\Omega_h(t)$, where for conciseness we use 
$t$ to stand for $B_\mathrm{thresh}$.
Then, $\Omega_h(t) = \Omega_d(t) + \Omega_m(t)$, where $\Omega_d$
is the portion taken from data, and $\Omega_m$ is the portion
taken from the model.  Another way of decomposing $\Omega_h$ is
into a true solid angle and error terms, i.e.,
$\Omega_h=\Omega + e_d(t) + e_m(t)$, where $e_d(t)$
is the error in the data portion for threshold $t$, and $e_m(t)$ is
the error in the model portion.  The model error can be parametrized
as a fraction of the model solid angle, such that $e_m(t)=a\Omega_m(t)$.
If $e_m$ is uncorrelated with $e_d$, then the fractional variance
in $\Omega_h$ is
\begin{equation}
  \label{eq:saerror}
  \var\{\Omega_h(t)/\Omega\} = \var\{e_d(t)/\Omega\} +
  \var\{a\}(\Omega_m(t)/\Omega)^2,
\end{equation}
and the hybridization
threshold is chosen to be the value minimizing this variance.

Essentially, the variable $a$ in the above discussion is a scaling
error that is common to all of the model points incorporated in the
hybrid beam.  A conservative method of estimating systematic
error is to assume that it is of the same order as the quantity
estimated.  In the above formulation, we represent this estimate 
by setting $a=1$.

\begin{figure}
  \begin{center}
    \includegraphics[width=5in]{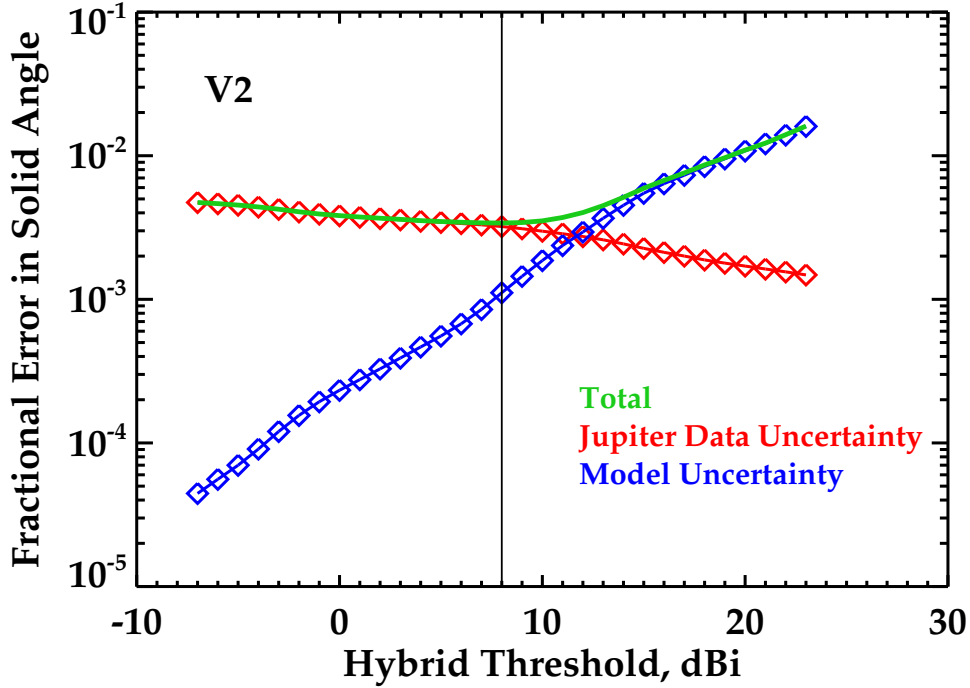}
    \caption{Fractional error in hybrid beam solid angle as a function
      of hybridization threshold, for the V2 DA.  As the
      hybridization threshold is raised, noisy Jupiter data are excluded,
      so that the Jupiter data uncertainty (\emph{red}) falls.  At the
      same time, the model uncertainty (\emph{blue}), estimated as a
      scaling error of $100\%$, increases because more of the
      two-dimensional model beam is used.  These contrary slopes produce a
      well-defined minimum in the total error (\emph{green}).
      Hybridization threshold values for the five-year analysis are chosen
      near the location of this minimum, as shown by the vertical line. 
      Plots for other DAs are similar.  The adopted thresholds
      are rounded to an even dBi unit and are constant for each
      frequency band, as shown in Table \ref{tab:transrad}.
      \label{fig:hybthresh}}
  \end{center}
\end{figure}

Figure \ref{fig:hybthresh} shows the fractional error in the hybrid beam solid
angle for the V2 beam on the A side, as a function of
$B_\mathrm{thresh}$.  To avoid a selection bias, $B_\mathrm{thresh}$ is
referred to the model rather than the data.  The errors contributed by
the data portion and the model portion are shown along with the total
error, which is computed using Eq. \ref{eq:saerror} with $a=1$.  For
the data, the solid-angle error is obtained as a function of
$B_\mathrm{thresh}$ from 100 Monte Carlo simulations in which model
input beams are combined with white noise appropriate to the Jupiter
data for each DA.
The contribution of the data portion increases with lowered threshold
as more of the noisy data are included in forming the hybrid beam. 
Conversely, the contribution of the tail portion decreases with lowered
threshold as less of the model is included.  The adopted
$B_\mathrm{thresh}$ values for the five-year analysis are obtained from the
locations of minimum total error in similar plots
made for all the A- and B-side beams.
These values are shown in Table
\ref{tab:transrad} together with the three-year equivalents.  The
five-year thresholds are lower than the three-year thresholds by
some $5-10$ dB, depending on DA.  Thus, we use significantly more
of the data than we have in the past.
\subsection{Sidelobes\label{sec:sl}}
For the five-year analysis, changes have been made in the sidelobe
sensitivity patterns that are distributed as part of the data release.
These patterns are in linear units of gain relative to isotropic, and
they are assembled from several types of data, including the Moon
observations described above.  In previous analyses, the conversion of
observed Moon brightness to gain assumed a temperature of 175K for the
Moon.  However, in K band the resulting gains were divided by 1.3 to
match adjacent pieces of the sidelobe pattern that were derived from
other data \citep{barnes/etal:2003}.   

For the five-year analysis, the calibration has been improved by
integrating the COBE DMR model of lunar microwave emission
\citep{bennett/etal:1992a} over the spacecraft and Moon ephemerides.
This procedure lowers the gain values in the Moon-derived parts of the
sidelobe pattern by factors of 1.35, 1.39, 1.45, and 1.51 for
the Ka, Q, V, and W bands, respectively, while confirming the
earlier adopted calibration of K band.

Additionally, some ground-based beam measurements, used within $\sim
10\arcdeg$ of each boresight, have been replaced with the augmented
main beam models described above (\S\ref{sec:moonextrap}).  The
ground-based data were taken under ambient conditions in the
Goddard Electromagnetic Anechoic Chamber (GEMAC), where the primary
mirror distortions are different from those under flight conditions.
This replacement affects at most $\sim 0.5\%$ of the sky for any given
DA.  

In the sidelobe reponse patterns, the area inside the main beams is set
to zero.  This area is expanded for the five-year analysis
(\S\ref{sec:beamprof}).

The differential signals tabulated in the \wmap\ TOD archive are
corrected for sidelobe contamination.  The overall effect can be
summarized in one number for each DA, called the \emph{sidelobe
  recalibration factor}, which is the factor by which the correction
changes the instrumental gain (since the dipole is detected in the
sidelobes along with other sources).  A sidelobe recalibration factor
of unity means that the sidelobe response is zero.  In the three-year
analysis, these factors differed from unity by $0.3\%-1.5\%$
\citep{jarosik/etal:2007}, and for the five-year analysis, they differ
from unity by $0.05\%-1.4\%$ \citep{hinshaw/etal:prep}.  The decrease
in the sidelobe correction is caused by the increased main beam area
together with the lower calibration of the Moon data.
\subsection{Symmetrized Beam Profiles\label{sec:beamprof}}
If the \wmap\ beam patterns could be well sampled in flight over
$4\pi$ steradians, then the distinction between main beams and
sidelobes would be arbitrary.  However, the two regimes are measured
by different methods, they are treated differently in the beam analysis, and
they are applied differently in the \wmap\ data reduction, so
that some reasonable boundary needs to be drawn.  We do so by using
the beam models for $\kmax=24$ to define a transition radius centered
on each boresight.  With the fitted mirror distortions, a separate
\dadra\ computation is done to extend each beam model into a wide
angular field, $11\arcdeg-13\arcdeg$ on a side.  Cumulative beam solid
angle is computed as a function of radius, and the radius containing
$99.9\%$ of the solid angle in the model is determined.  The
transition radius is then fixed at a round number encompassing the
computed radii for both the A and the B sides.  The adopted values are
$7\fdg0$, $5\fdg5$, $5\fdg0$, $4\fdg0$, and $3\fdg5$ for the bands K,
Ka, Q, V, and W, respectively.  Compared to the one-year and
three-year analyses, the transition radius is increased, as
shown in Table \ref{tab:transrad}.

This expansion of the main beam region has the useful consequence of
mitigating the sidelobe correction.  However, the main
beam now includes lower $S/N$ observations, to which the
main beam solid angle is sensitive.  Similarly, the profile-fitting
algorithms of the first- and third-year analyses can no longer be used
as previously implemented, because the fitting function is difficult
to constrain over the entirety of the new main-beam radius.

In the one- and three-year analyses of azimuthally symmetrized beams,
the radial profiles were modeled with basis functions of the form
\begin{equation} 
  H_{2n}\left(\frac{\theta}{\sigma_h}\right)
  \exp\left(-\frac{\theta^2}{2\sigma_h^2}\right), 
\end{equation} 
where $\theta$ is the angle from the beam center, $H_{2n}(x)$ is a
Hermite polynomial of even order, and $\sigma_h$ determines the width
of the Gaussian.  The first of these basis functions is a pure
Gaussian, which is a good fit to the main lobe of the beam, both
theoretically and in reality.  The other basis functions parameterize
the deviations from Gaussianity.  The Hermite fit is limited to the
well-characterized part of the beam, within a given cutoff angle
$\theta_c$ of the beam peak.  In the five-year analysis, $\theta_c$ is
increased as compared to the three-year value, because of improvements
in the data and analysis described above.  However, this presents two
problems with the Hermite polynomials.

First, the basis functions do not extend far enough in $\theta$.
Since the Hermite polynomial $H_{2n}(x)$ is a polynomial of
order $2n$, and the function $x^{2n}\exp(-x^2/2)$ has peaks at
$x=\pm\sqrt{2n}$, the basis functions of order $2n$ extend 
to $\theta \approx \sigma_h \sqrt{2n}$, beyond which, the function is
exponentially suppressed.  Because this angle increases only with
the square root of the order, many basis functions are required
to cover the required domain, e.g., $\theta
\lesssim 40\sigma_h$ in W band.  Second, numerical problems arise in
computing the Hermite polynomials of higher order than $\sim 150$.
The combination of these two problems rules out the use
of Hermite functions in the five-year analysis.

The use of a fitting basis provides a smooth fit through noisy
portions of data, and also provides a convenient mechanism for the
derivation of a beam covariance matrix via the formal statistical
errors in the fit. Because of these benefits, a number of possible sets of
basis functions have been explored for the five-year beam data
using simulations.

Beam profile simulations test the accuracy to which various sets of
basis functions reproduce the known input beams and window functions.
A variety of noisy simulated beams is constructed, then fitted.  The
simulations include pure \dadra\ models as well as hybrids of two
\dadra\ models.  In the case of hybrids, one beam model with noise
added is used to represent the Jupiter data, and another model without
noise is used for the beam tails.  The hybridization thresholds
(\S\ref{sec:hybrid}) are also varied, as is the overall scaling of the
beam tails.  The result of this testing is that ultimately, no one set
of basis functions recovers the input beam solid angle and window
function.  One of the impediments seems to be the nature of the
five-year hybrid itself, which is noisy at intermediate angular scales
within the transition radius.  Functional fit residuals in that region
typically cause a bias of $\sim 0.5\%$ in the recovered solid angle. 

Owing to this difficulty, the method of basis function fitting is not
used in the five-year analysis.  The adopted hybrid beam profiles are
left in the radially binned form, in spite of the noise that remains at
low gain levels.  Simulations show better recovery of solid angle from
the resulting beams than from any of the attempted basis function fits.

The applicability of symmetrized beam profiles depends on the degree
to which the assumption of azimuthal beam symmetry is justified.  The
\wmap\ scan strategy mitigates the effect of noncircularity in the
beams by sampling most sky pixels over a wide range of azimuth angles.
The effects of residual noncircularity are of potential importance for
CMB power spectrum analysis primarily in Q band at $\ell \gtrsim 500$,
where the effect can reach several percent in $C_\ell$; however, Q-band
data have low statistical weight in this $\ell$ range and are not
used in the TT power spectrum analysis \citep{nolta/etal:prep}.  Mathematical
details, together with plots of the relevant correction to $w_\ell$,
are given in Appendix B of \citet{hinshaw/etal:2007}.
\subsection{Beam Transfer Functions and Errors\label{sec:btf}}
To compute beam transfer functions, the radial distance, $\theta$,
from the beam centroid is computed for each of the data points in the
two-dimensional A- and B-side hybrid beams.  Radial profiles
$b^S(\theta)$ are constructed by sorting all hybrid data into equally
spaced bins of $\Delta\theta=0\farcm25$ in width, and taking the mean of
each bin.  The radial profile only extends out to the transition
radius.  The beam transfer functions are evaluated using the Legendre
transform: 
\begin{equation} 
  B_\ell = \Omega_B b_\ell = 2 \pi \int b^S(\theta) P_\ell(cos\,\theta)
  d \cos\theta.\label{eq:legendre} 
\end{equation} 
Numerically, the
integration is performed by summing over rectangular bins of
$\Delta\theta=0\farcm25$.

As described above (\S\ref{sec:moonextrap}), the sidelobe data of the
Moon motivate an attempt to augment the fitted distortions of the
primary mirror ($\kmax=24$) with random distortions that are
extrapolated to finer spatial scales, i.e., $\kmax=250$.  These added
distortions affect the hybridized beam through their effect on the
outermost, low-gain part of the main beam model.  One way of
testing the effect of systematic error in the modeled beam tail is to
rescale the extrapolated distortion amplitudes up or down as a group,
with $100\%$ correlation.  The resulting distortion amplitudes
are used to compute new model beams, which are processed through
hybridization with flight data and transformation to $b_\ell$.

\begin{figure}
  \begin{center}
    \includegraphics[height=6.5in]{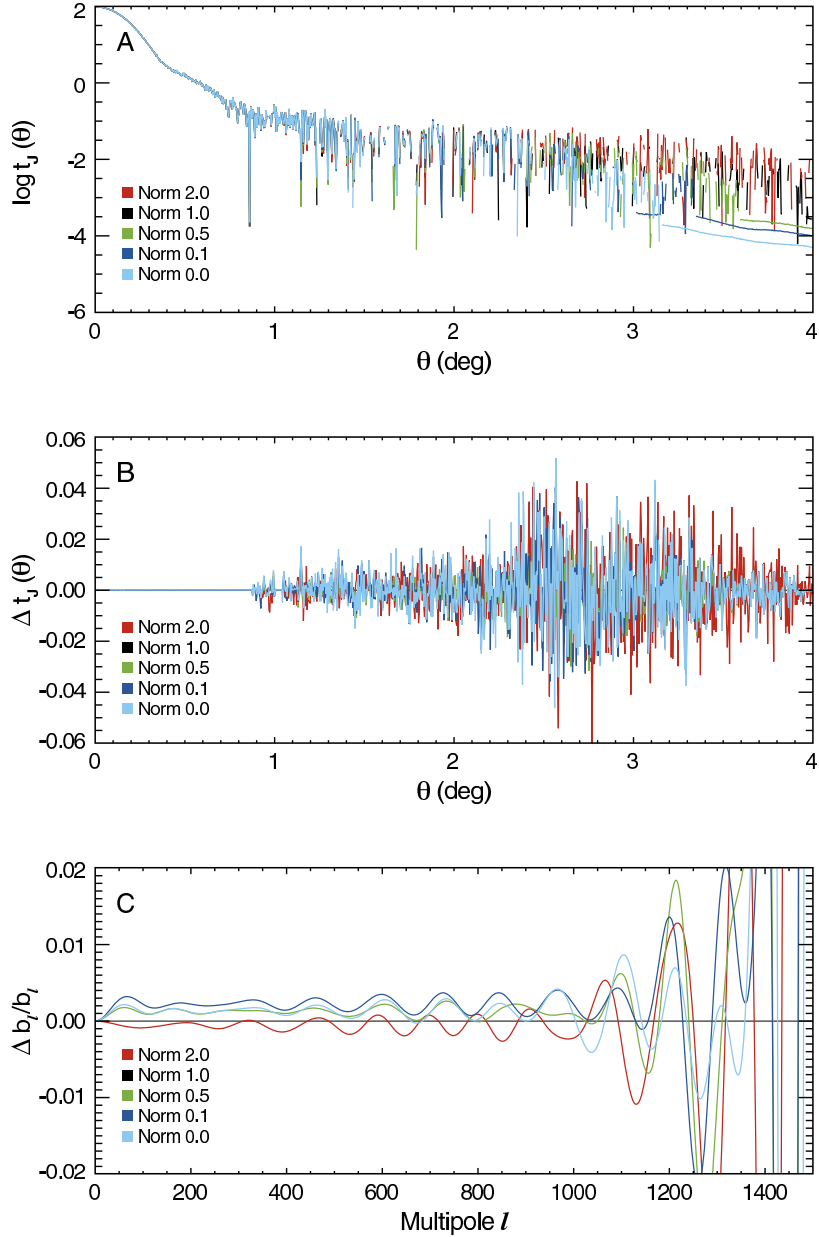}
    \caption{(Top)  Symmetrized radial profiles of hybridized, binned
      flight beams for the V2 DA.  The central, high $S/N$ part
      of the beam is taken directly from flight data of the planet Jupiter,
      whereas the part of the beam below a given gain cutoff is taken from
      beam models.  The set of beam models shown comprises several normalizations
      of the extrapolated primary-mirror distortions; the
      normalization favored in the analysis is zero, meaning that the extrapolated
      distortions are omitted.  The noise-free lines at radii
      $3\arcdeg-4\arcdeg$ are portions of the lower-normalization
      profiles that include model points only.  (Middle) Same profiles
      as in the top panel, after subtraction of the one with
      normalization 1.0.  (Bottom) Beam transfer functions corresponding to
      the depicted beam profiles, relative to the one with normalization 1.0.  
      Cf.  Figure \ref{fig:noiseless},
      especially the bottom panel.  The beam transfer functions at
      $\ell\gtrsim100$ are close to what is expected from the noise-free
      simulations, implying good solid angle recovery.  \label{fig:flthyb}}
  \end{center}
\end{figure}

\begin{figure}
  \begin{center}
    \includegraphics[height=6.5in]{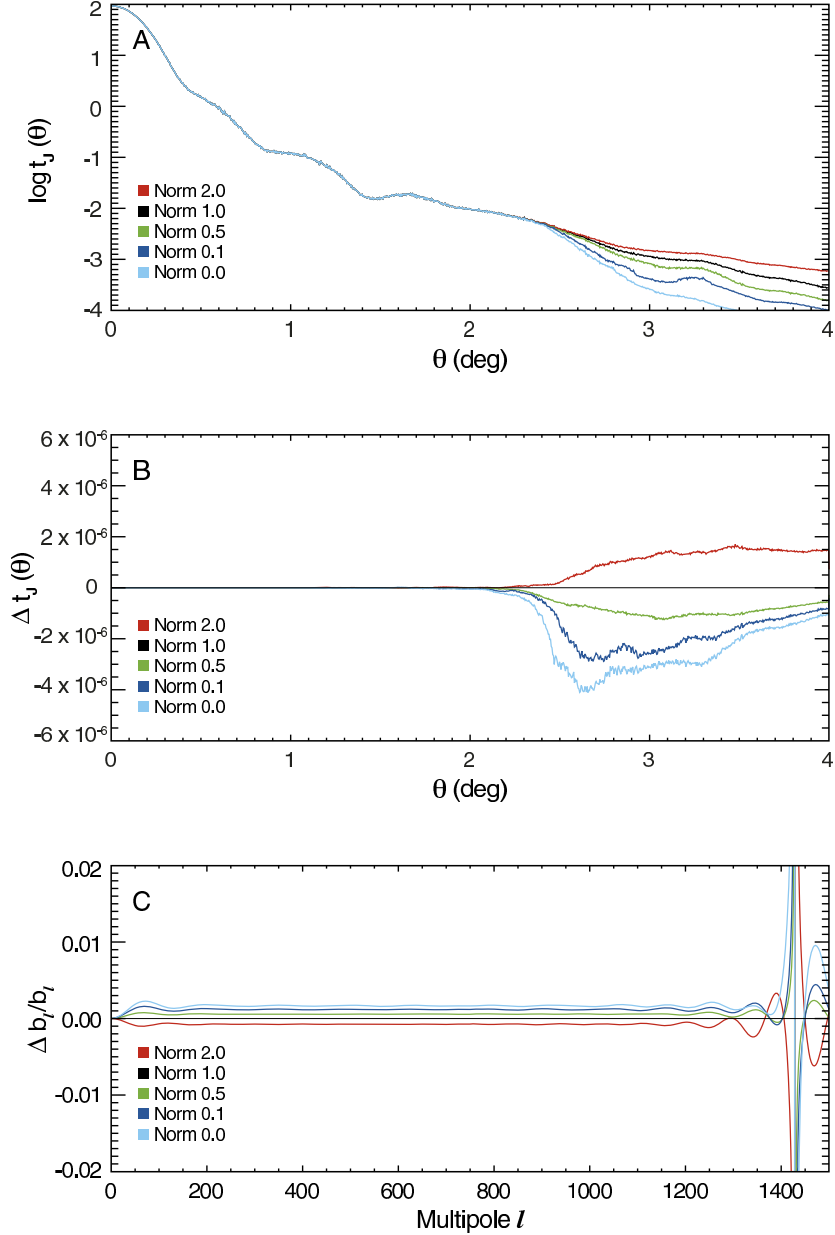}
    \caption{(Top)  Radially binned noise-free simulations of
      hybridized beams for the V2 DA.  The central part of the beam is
      taken from the adopted \dadra\ model without augmented
      distortions.  The part of the beam below a given gain cutoff is
      taken from augmented beam models with various normalizations of
      the extrapolated primary-mirror distortions.  The analysis of
      hybrid beams including real Jupiter data favors the zero
      normalization, meaning that the extrapolated distortions are
      omitted.  (Middle)  Same profiles as in the top panel, after
      subtraction of the one with normalization 1.0.
      (Bottom) Beam transfer functions corresponding to the depicted
      beam profiles, relative to the one with normalization 1.0.  
      Fits using flight hybrid beams should
      approximate the curves shown here. 
      \label{fig:noiseless}}
  \end{center}
\end{figure}

The effect of this type of distortion rescaling on the beam transfer
functions is shown in the bottom panel of Figure \ref{fig:flthyb} for
flight data, and Figure \ref{fig:noiseless} for a noiseless
simulation.  Rescaling the added mirror distortions changes the slope
of $b_\ell$ between $\ell=0$ and $\ell\sim 100$, while shifting
$b_\ell$ up and down for $\ell\gtrsim100$.  Scale factors in the range
$0-2$ result in a $\sim 0.3\%$ total range of variation in the
high-$\ell$ value of $b_\ell$.

\begin{figure}
  \begin{center}
    \includegraphics[height=5.5in]{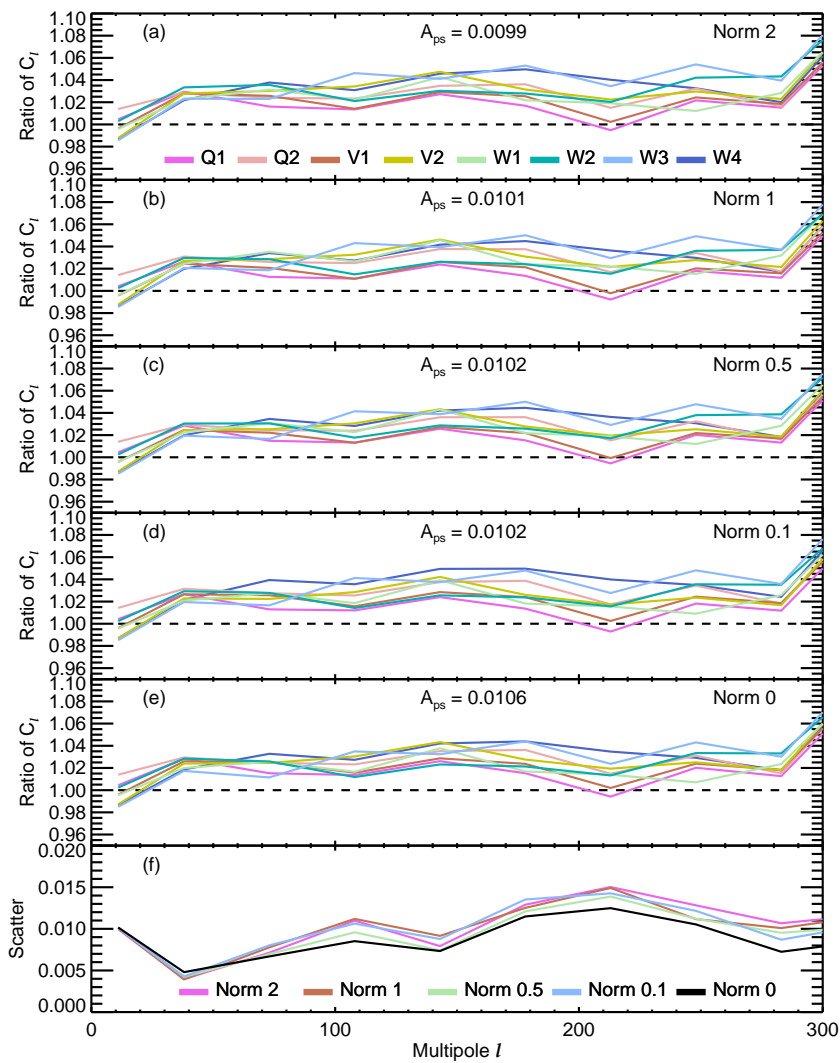}
  \end{center}
  \caption{Consistency of CMB power spectra across frequency bands,
    for window functions derived from various normalizations of the
    extrapolated primary mirror distortions.  Complete omission of such
    extrapolated distortions from the main beam model is justified by
    this criterion.  The CMB spectra and unresolved point-source coefficients
    ($A_\mathrm{ps}$) in
    this plot are from a preliminary stage of analysis and are not the
    final five-year \wmap\ results.  (a)-(e): Mean of year-to-year cross
    spectra in each DA, relative to the final combined power spectrum
    from the three-year analysis.  The applied $w_\ell$ are derived from
    hybridized beams in which the tail is from a beam model with
    extrapolated primary mirror distortions; hybridization thresholds in
    each DA optimize solid angle error for the nominal amplitude of
    these added distortions (Figure \ref{fig:ps-ab}).  Spectrum is binned
    in $\ell$ with a bin size of $\Delta\ell=35$.  The panels differ in
    the scaling of the extrapolated distortion amplitude on the mirror:
    (a), 2;  (b), 1; (c), 0.5; (d), 0.1;  (e) no extrapolated
    distortions.  (f)  Scatter among the DAs in each $\ell$ bin for the
    five normalizations of the extrapolated mirror distortions.
    Omitting the extrapolated distortions (Norm 0, \emph{black}) minimizes
    the scatter in the CMB power spectrum over most of this $\ell$
    range, which includes the first peak near $\ell\sim200$.
    \label{fig:multinorm}}
\end{figure}

These scalings of the added distortions have been tested for their
ultimate effect on CMB power spectra.  Figure \ref{fig:multinorm}
shows the results of this test for scale factors of $2$, $1$, $0.5$,
$0.1$, and $0$, respectively.  Each panel shows a mean of year-by-year
CMB cross power spectra computed from the five-year data set for each
of the 8 \wmap\ DAs Q1--W4.  The spectra are all computed using the
MASTER estimator, and they are corrected for $b_\ell$ derived from
augmented beams, characterized by distortion scale factors as
indicated.  For plotting, each such power spectrum is divided by the
final MASTER power spectrum from the three-year \wmap\ 
analysis\footnote{The CMB spectra and point source coefficients in
  this plot are from a preliminary stage of analysis and are not the
  final five-year \wmap\ results.}. In each case, a contribution
from unresolved point sources is fitted and removed.  In general, the
result is that lower values of the scale factor give better
consistency between microwave frequency bands for the CMB.  Indeed, on
this criterion, there is no clear reason to prefer a scale factor
greater than zero.

As a result, the way the extrapolated random-phase mirror distortions
are handled is by omitting them from the adopted beams and $b_\ell$,
while the actual fitted mirror distortions with $\kmax\leq24$ are
retained, via the model part of the hybrid beam.  However, we
incorporate into the error analysis an estimate of the systematic
error in the faint part of the model, by assuming that this error is
of the same order as the adopted model, just as we do for optimizing
$B_\mathrm{thresh}$ (\S\ref{sec:hybrid}).  Monte Carlo experiments
done on the primary mirror distortions suggest that this $100\%$
scaling error is conservative.

\begin{figure}
  \begin{center}
    \includegraphics[height=7in]{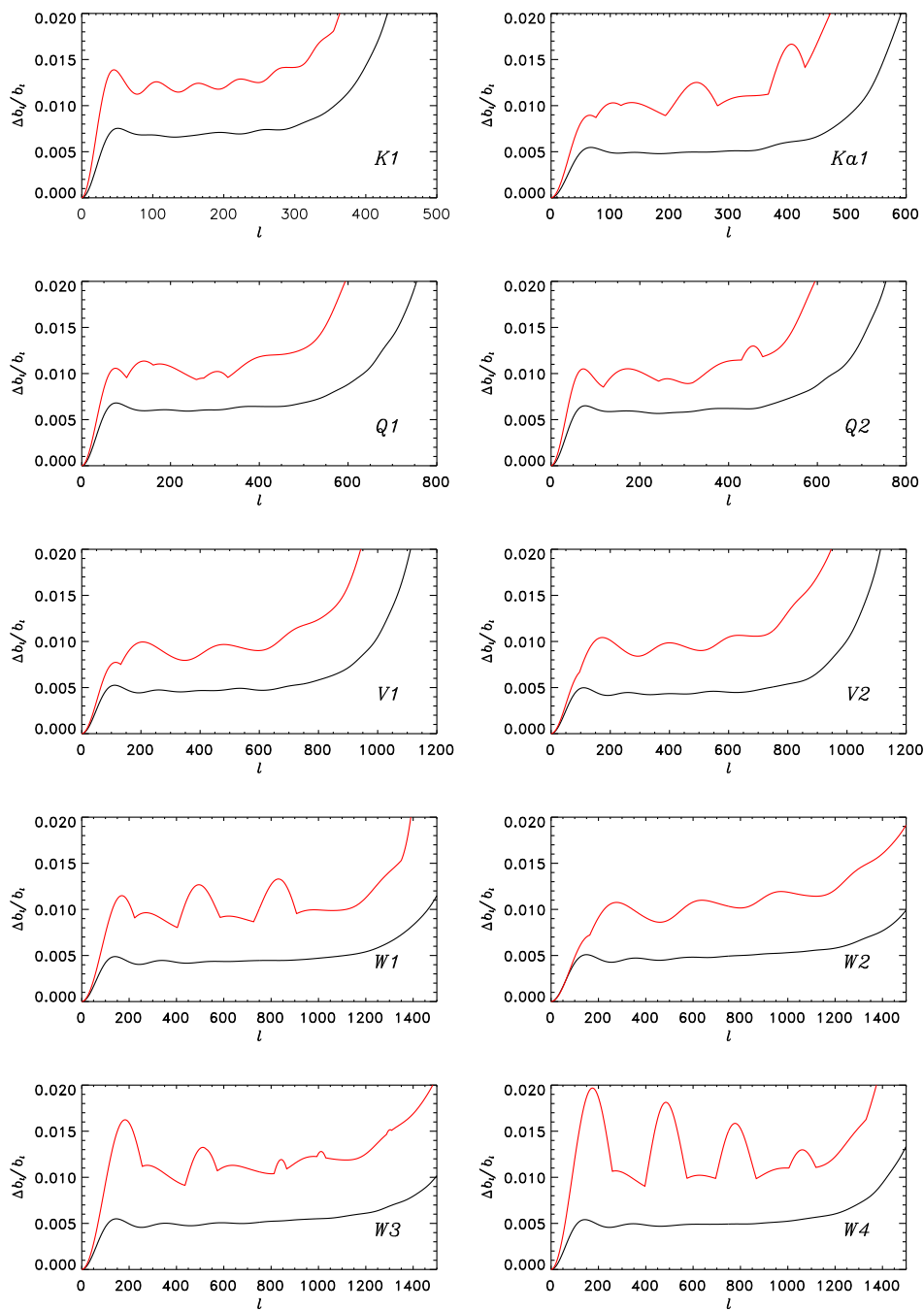}
    \caption{Relative error in beam transfer functions ($\Delta
      b_\ell/b_\ell$) for the five-year beams (black) vs. the three-year
      errors (red) \citep{jarosik/etal:2007}.  The five-year uncertainties
      are typically a factor of $\sim 2$ better than three-year
      uncertainties.\label{fig:errband}}
  \end{center}
\end{figure}

Combined errors in $b_\ell$, which arise both from observational scatter in the
Jupiter measurements and from the scaling error in the model,
are estimated using Monte Carlo simulations of the
hybridization.  The \dadra\ flight models are used to represent
the true input beams.  These models are sampled to match the observed
beam positions in the five-year flight archive.  Based on the chosen
hybrization threshold, white noise is added to the model for the
points that would be taken from Jupiter observations in the actual
analysis.  The model points that are substituted for the low-gain tail
are multiplied by a common, normally distributed scaling factor with
$\mu=\sigma=1$.  Because the beam modeling is expected to be common mode
across frequencies, the same scaling factor is used for every DA in a
given Monte Carlo realization.  A total of 5000 beam realizations is
computed, each comprising all ten DAs.  The beam transfer function is
computed for each beam realization as described above for flight data,
and the standard deviation of the realizations at each $\ell$ is used
as the diagonal of the covariance matrix.  Figure \ref{fig:errband}
shows the resulting error bands for each DA, compared to the adopted
errors from the three-year analysis.

The chosen radial bin size of $0\farcm25$ is the smallest width permitting
all the bins to be populated.   However, the bin sizes $0\farcm5$ and
$0\farcm75$ have also been tested, in order to ensure that window
function results are not affected.  For the $\ell$ range of interest,
the bin size contributes negligibly to the smearing of the beam, even
in W band, and has no effect on the estimated error bars.

\begin{figure}
  \begin{center}
    \includegraphics[width=5in]{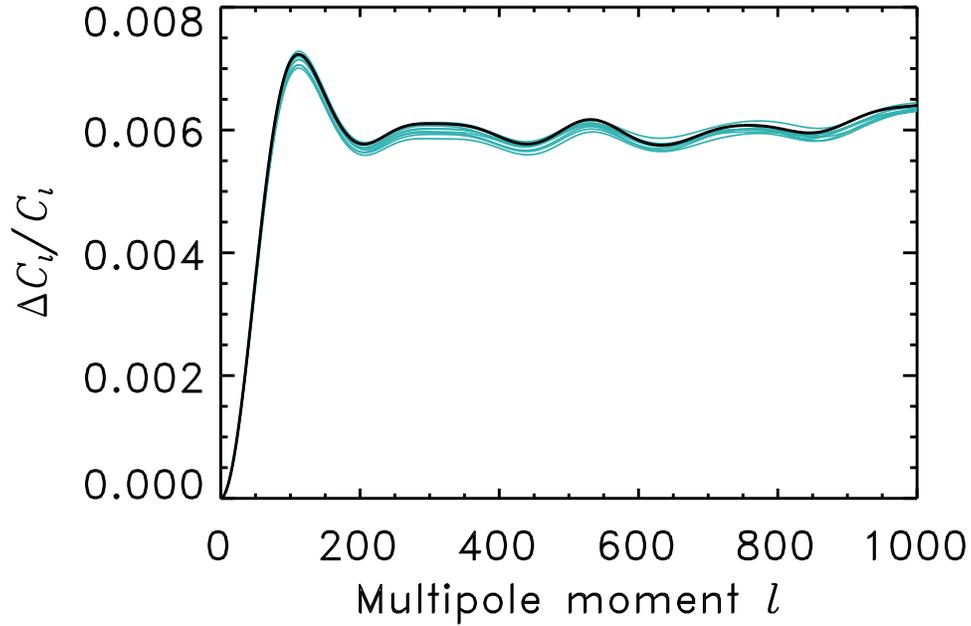}
    \caption{Relative error component that is due to estimated beam errors in 
      the final \wmap\ TT power spectrum, which is
      combined from V and W band data.  Cyan:  eight independent instances
      of the square root of the diagonal of the covariance matrix for the
      coadded VW $C_\ell$; each instance is based on 5000 Monte Carlo
      realizations of V and W beam errors.  Black:  the instance that has
      been chosen for the $C_\ell$ error bar, because it is approximately the
      upper envelope.
      \label{fig:vwwinerr}}
  \end{center}
\end{figure}

Monte Carlo realizations of $b_\ell$ for each DA are used in estimating
the full covariance matrix of the coadded TT power spectrum in V and W. 
For CMB analysis, the error in the error is important.  From several
independent Monte Carlo runs of 5000 realizations apiece, the combined
VW window function covariance has an error of $\sim \pm3\%$ in the
diagonal elements \citep{hinshaw/etal:prep}.  The beam-related
errors in the coadded TT power spectrum are shown as a function of
$\ell$ in Figure \ref{fig:vwwinerr}.

\begin{figure}
  \begin{center}
    \includegraphics[height=6.7in]{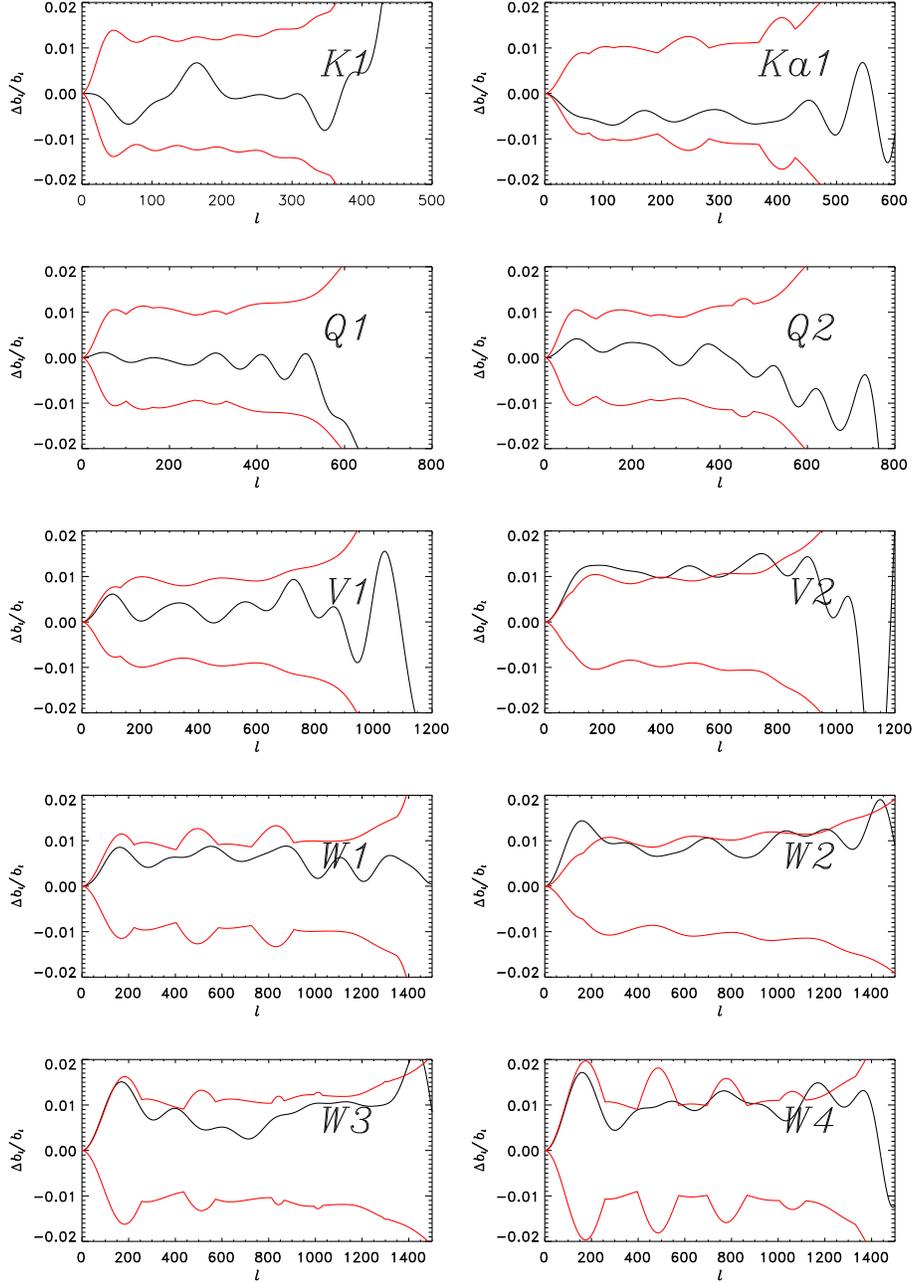}
    \caption{Consistency within $\sim 1\sigma$ of five-year beam 
      transfer functions with three-year beam transfer functions.
      \emph{Black}: difference of three-year minus five-year $b_\ell$,
      relative to the five-year $b_\ell$;  the low-$\ell$ rise or fall
      for several DAs reflects solid angle changes detailed in Figure
      \ref{fig:saincrease}.  \emph{Red}:  three-year $1\sigma$ errors.
      For this plot, the beam profiles used to compute $b_\ell$ are
      extended by including the profile of the inner portion of each
      sidelobe response pattern, and the resulting composite profiles
      are integrated out to the five-year transition radii.  This
      removes the effect of the larger five-year transition radius.
      \label{fig:p2errband}}
  \end{center}
\end{figure}

The comparison of three-year and five-year beam transfer functions is
shown in Figure \ref{fig:p2errband}.  Plotted is the relative change in
$b_\ell$, i.e., $(b_\ell^\mathrm{3yr} - b_\ell^\mathrm{5yr})/
b_\ell^\mathrm{3yr}$.  Each beam is integrated out to the five-year
transition radius; for this plot, the three-year beams are extended
using the three-year sidelobe response patterns.  Also plotted are $\pm
1\sigma$ errors of $b_\ell^\mathrm{3yr}$.  The three-year and
five-year $b_\ell$ are consistent, with $\sim 1\sigma$ changes in V2 and
W1--W4.  For these DAs, the change has the form of a plateau for
$\ell\gtrsim 200$, reflecting an increase in the main beam solid angle
for five years.  This increase raises $B_0$ relative to $B_{200}$.  The
differences plotted in Figure \ref{fig:p2errband} are taken with a
sign convention reflecting the difference in the final power spectrum. 
Thus $\Delta b/b_\ell = 1\%$ implies a $2\%$ change in $w_\ell$, in the
sense that the high-$\ell$ power spectrum increases by $\sim 2\%$.

\begin{figure}
  \begin{center}
    \includegraphics[height=6in]{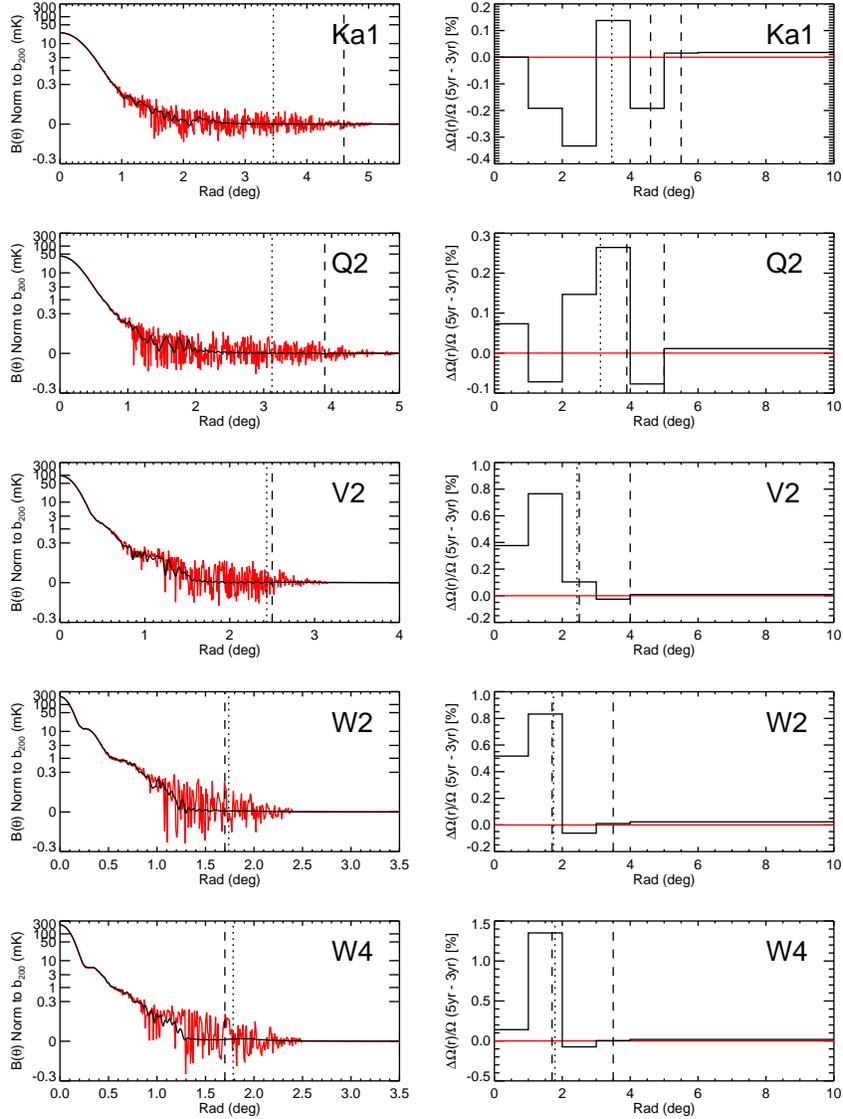}
    \caption{Much of the solid angle change between the three-year
      and five-year beams arises inside the three-year main-beam
      boundaries.   In this figure, the beam profiles are extended to 
      a radius of $10\arcdeg$ using the three-year or five-year sidelobe 
      response pattern, respectively, and the beams are normalized to give 
      the same $b_{200}$ for both three years and five years.  Solid 
      angle changes by $|\Delta\Omega_B|/\Omega_B<0.5\%$ for the K1, 
      Ka1, Q1, Q2, and V1 DAs, and by $0.8\%\leq\Delta\Omega_B/\Omega_B\leq1.5\%$ for
      the V2 and W1--W4 DAs.  \emph{Left}:  Five-year symmetrized hybrid beam
      profiles (\emph{red}) and three-year Hermite-fitted beam
      profiles (\emph{black}) for selected DAs. The five-year profiles
      include Jupiter data and so are noisy, whereas the three-year
      profiles are the functional fit only. Dashed line:  three-year transition radius (Table
      \ref{tab:transrad}).  Dotted line: radius where five-year hybrid
      beams consist of $50\%$ data and $50\%$ model.  \emph{Right}:
      Change in beam solid angle from the three- to the five-year
      analysis, as a function of radius, in annuli of $1\arcdeg$,
      expressed as a percentage of the five-year $\Omega_B$.  Dashed lines:  
      Transition radii for three years and five years,
      respectively.  Dotted lines: $50\%$ data radius of hybrids, as
      in left column.
      \label{fig:saincrease}}
  \end{center}
\end{figure}

Solid angle changes are necessarily attributable to changes in the
symmetrized beam profiles.  Selected beam profiles are compared in the
left column of Figure \ref{fig:saincrease}.  The right column
partitions the solid angle difference between five years
($\Omega_B^\mathrm{5yr}$) and three years ($\Omega_B^\mathrm{3yr}$)
into $1\arcdeg$ radial bins.  The contribution of each bin is plotted
as a percentage of $\Omega_B^\mathrm{5yr}$.  In this case, the beam
profiles are extended to a radius of $10\arcdeg$ using the far
sidelobe patterns.  Also, solid angles are normalized in such a way as
to equalize $b_{200}$ between the resulting three-year and five-year
beam transfer functions.  Most of the beam profile change, and
therefore most of the solid angle change, is just inside the
three-year transition radius (dashed line).  Compare with Figure
\ref{fig:kmx}, which shows, for the pure model case, how the increase
in the fitted $\kmax$ of the mirror distortions for five years of data
increases the solid angle inside the three-year radius.

In summary, the solid angle increase appears to result primarily from
the improved beam modeling, together with the extension of the main
beam treatment to larger radii, both resulting from the increased
$S/N$ of the Jupiter data.  Optimally hybridized two-dimensional beams
are symmetrized and reduced to radial profiles in an unbiased way by
averaging in annuli, and the resulting profiles are transformed
directly to $b_\ell$.
\section{Radiometry of Planets Useful for Calibration\label{sec:src}}
\subsection{Jupiter}
We adopt the analysis approach described in \citet{page/etal:2003e}
for reduction of the planet observations.  The 
\wmap\ full-sky maps exclude
observations made with Mars, Jupiter, Saturn, Uranus, or Neptune near
any main beam boresight, with an exclusion radius of $1\fdg5$.
In turn, the sky maps are used to remove the background sky signal from
planet data.  Since the solid angle of each planet is much less than
that of the \wmap\ beams, a beam map is built up by binning
observed antenna temperatures for a planet in a focal plane coordinate
system.  Rather than being normalized, this beam map may be left in 
antenna temperature and Legendre transformed (Eq. \ref{eq:legendre})
to produce an unnormalized beam transfer function, 
$T_\mathrm{J}^m B_\ell$, where $T_\mathrm{J}^m$ is the peak antenna temperature
of Jupiter.  But $B_0$ is the beam solid angle
$\Omega_B$, so that $T_\mathrm{J}^m B_0 = T_\mathrm{J}^m \Omega_B$ 
\citep{page/etal:2003e}, and
$T_\mathrm{J}^m B_0/\Omega_\mathrm{J}^\mathrm{ref} = T_\mathrm{J}$,
where $T_\mathrm{J}$ is the brightness temperature of the Jupiter
disk, and $\Omega_\mathrm{J}^\mathrm{ref}$ is the fiducial
solid angle $2.481\times 10^{-8}$ sr for a Jupiter--\wmap\ distance
of 5.2 AU \citep{griffin/etal:1986}.  The error in $T_\mathrm{J}$ is then the sum in quadrature
of the error in $\Omega_B$ with the estimated \wmap\ gain calibration
error of $0.2\%$ \citep{hinshaw/etal:prep}.  The results of this
procedure are given in Table \ref{tab:juptemp} for five years of
Jupiter data in each DA.

The main difference between the Legendre transform method and a direct
integration of the two-dimensional beam map is that the Legendre
transform uses a symmetrized beam profile.  Integration of the beam
map yields solid angles within the errors of the above approach.  The
results in Table~\ref{tab:juptemp} are consistent with the band
averaged ones previously reported by \citet[\S2.4]{page/etal:2003e}.
Currently, the error in $T_\mathrm{J}$ is limited by the $0.5\%$ error
in beam solid angle and the $0.2\%$ gain uncertainty.  

Season-by-season radiometry of Jupiter is given in Table \ref{tab:juptime}.
The values are computed using a template-fitting technique.  Radial
profiles are produced for each DA for each Jupiter season, then fitted
linearly against the mean five-year Jupiter radial profile.  Season 2
is omitted because Jupiter is approaching the Galactic anticenter,
making background subtraction problematic.  Our data place an upper
limit on the time variability of $T_\mathrm{J}$ as a function of
orbital phase of $0.3\% \pm 0.5\%$.  We conclude that our radiometric
observations are consistent with the absence of variability in the Jupiter
brightness temperature at this level.

In view of the stability and low errors of these measurements, Jupiter
radiometry is the preferred method of transferring the \wmap\ dipole
calibration to another microwave instrument.  The key requirement for
such an effort is knowledge of the beams.  The error values given in
Table \ref{tab:juptemp} include \wmap\ beam errors via error in solid
angle, as well as the fundamental gain uncertainty relative to the
dipole.
\subsection{Other Planetary Calibrators}
Millimeter-wave brightnesses of other planets are also of
potential interest as calibrators.  For example, for the \wmap\ W band
beams ($3.2$ mm), peak antenna temperatures of $\sim 200$ mK, $\sim 35$
mK, and $\sim 6$ mK are produced by Jupiter, Saturn, and Mars,
respectively.  A preliminary analysis of the \wmap\ five-year Mars and
Saturn observations has been undertaken.

\begin{figure}
  \begin{center}
    \includegraphics{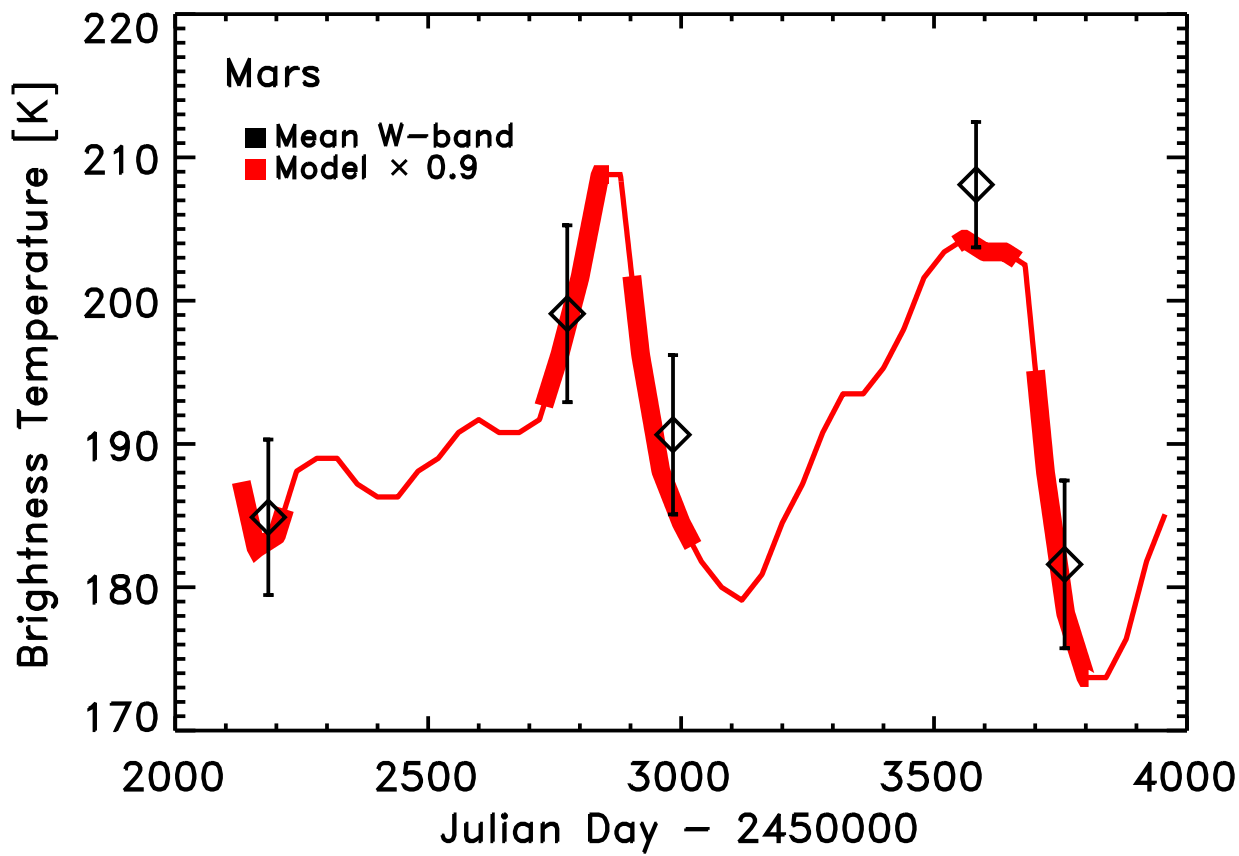}
    \caption{Comparison of \wmap\ observations (Table \ref{tab:satmars}) 
      to the Mars brightness model of \citet{wright:1976,wright:2007},
      evaluated at a wavelength of $3.2$ mm (W band).  Mean \wmap\
      meaurements are shown for each observing season
      (\emph{diamonds}), with error bars indicating the scatter among
      \wmap\ DAs W1--W4.   Model values (\emph{red}) are rescaled by
      $0.9$ to bring them into overall agreement with the
      measurements; thick portions of the line indicate observing
      seasons.  \wmap\ data are referenced to a fiducial distance of
      $1.5$ AU and a solid angle of $\Omega_\mathrm{Mars}^\mathrm{ref}
      = 7.156 \times 10^{-10}$ sr \citep{hildebrand/etal:1985}.  There
      are significant variations in the observed brightness
      temperature due to both geometric and physical factors, and
      thus, some care must be exercised before taking Mars as a
      calibration source.  \label{fig:mars}}
  \end{center}
\end{figure}

Mars is attractive as a calibration source because it is relatively
bright.  However, significant variations in the observed brightness
temperature can occur because of the viewing geometry.  Moreover, the
radiating properties of the inhomogeneous, pitted planetary surface
complicate the determination of an appropriate reference brightness.
Figure \ref{fig:mars} shows a thermal model developed for the infrared
by \citet{wright:1976,wright:2007}.  The model is evaluated at $3.2$
mm (W band) as a function of the time within the \wmap\ five-year
timeline, which includes five Mars observing seasons (fewer than for
Jupiter because of the relative orbital velocity of Mars).  The
predicted variation in brightness temperature over an observing season
can be as much as $\sim 20$ K.  The mean $3.2$ mm temperature and the
scatter among the four W DAs are also shown 
for \wmap\ data binned by Mars observing season (Table \ref{tab:satmars}).  
The model is higher than the \wmap\ measurements by $\sim 10\%$, so
that a renormalization factor of $0.9$ is applied to the model in the
plot.  We use this Mars model partly because of its previous use for
Earth-based infrared calibration and the convenient availability of
the code \citep{wright:2007};  for a model including the effects of a
dusty atmosphere and polar caps, see \citet{simpson/etal:1981}.  The
Mars data are referenced to a fiducial distance of $1.5$ AU and a
solid angle of $\Omega_\mathrm{Mars}^\mathrm{ref} = 7.156 \times
10^{-10}$ sr \citep{hildebrand/etal:1985}.

\begin{figure}
  \begin{center}
    \includegraphics{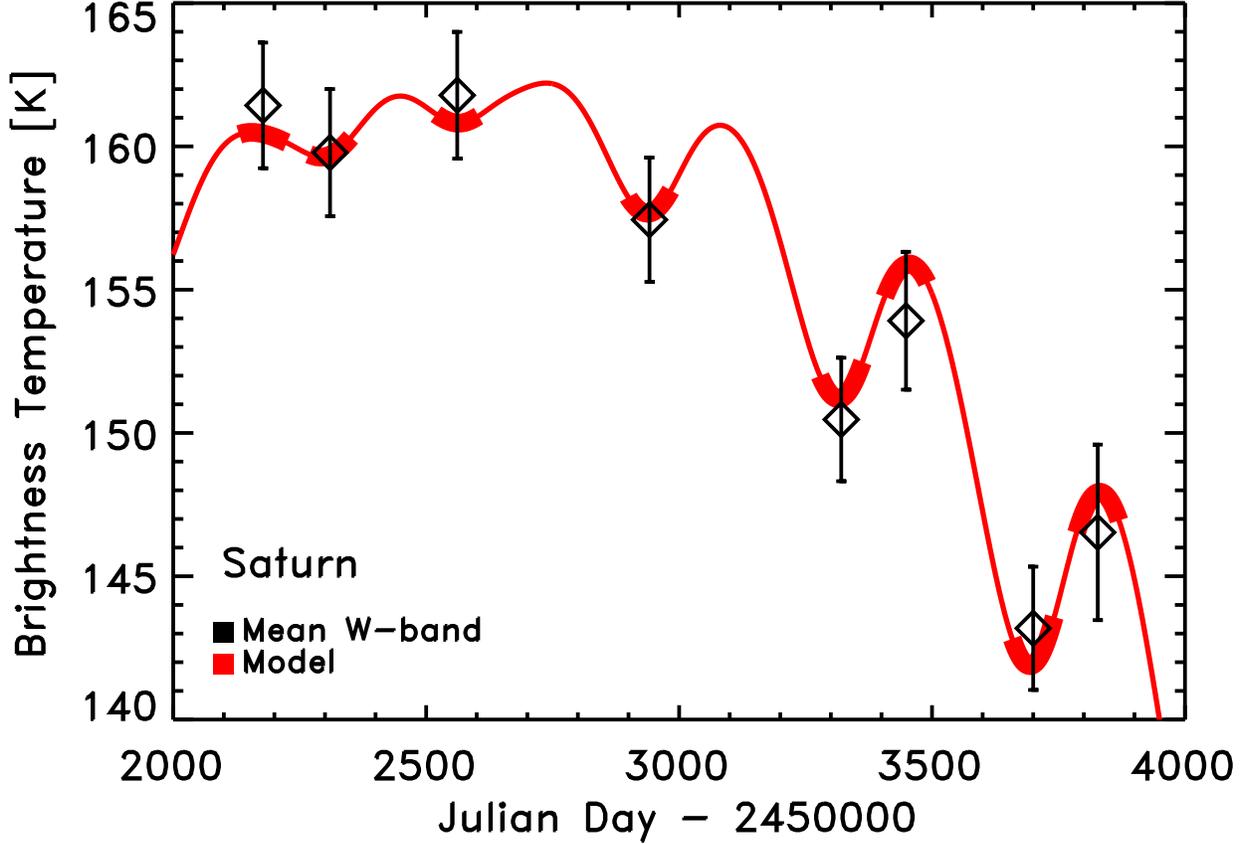}
    \caption{Season-by-season \wmap\ radiometry of Saturn in the W band
      (Table \ref{tab:satmars}).
      \emph{Diamonds}:  Mean \wmap\ measurements for each of eight
      observing seasons, with error bars indicating the scatter among
      \wmap\ DAs W1--W4.   \wmap\ data
      are referenced to a fiducial distance of $9.5$ AU, corresponding
      to a Saturn solid angle of $\Omega_\mathrm{Sat}^\mathrm{ref} =
      5.101 \times 10^{-9}$ sr \citep{hildebrand/etal:1985}.
      \emph{Red line}:  simple fitting
      model of the form $T_\mathrm{Sat} = T_0 + \alpha\sin i$, where $i$
      is the inclination of the ring plane from our line of sight.
      Thick portions of the line indicate observing seasons.
      Fitted parameters are $\alpha=-132\pm 16$ and $T_0=102 \pm 7$. 
      Although the model appears to capture geometric aspects of the
      observations surprisingly well, it lacks the physical underpinning
      to be used predictively.
      \label{fig:saturn}}
  \end{center}
\end{figure}

Saturn's apparent brightness is even greater than that of Mars, but
the theoretical understanding of the radiometry is less developed
\citep{ulich:1981,epstein/etal:1980,hildebrand/etal:1985}.  A special
consideration is Saturn's ring system, of which the viewing aspect
from the Earth changes over the course of Saturn's 29 year orbital
period.  In Figure \ref{fig:saturn}, mean seasonal W-band brightness
temperatures as measured by \wmap\ (Table \ref{tab:satmars}) are 
shown as black diamonds.  These data show a clear decrease
in observed temperature with time, a trend which correlates extremely
well with the decreased viewing cross-section of Saturn's rings in
the same time interval.  A simple model of the form $T_\mathrm{Sat} =
T_0 + \alpha\sin i$, where $i$ is the inclination of the ring plane
from our line of sight, is fitted to the data and plotted in red.  The
fit results are $\alpha=-132\pm 16$ and $T_0=102 \pm 7$.  Possible
physical causes for the temperature decrease include the decreasing
projected radiating area of the rings, a less favorable viewing angle
for the ``hot spot'' at the south pole of Saturn, and Saturn's
oblateness.  These causes will be the subject of future investigation.
The Saturn data are referenced to a fiducial distance of $9.5$ AU,
corresponding to a Saturn solid angle of
$\Omega_\mathrm{Sat}^\mathrm{ref} = 5.101 \times 10^{-9}$ sr
\citep{hildebrand/etal:1985}.

Clearly, Jupiter remains the only \wmap\ source that can be
recommended as an instrument calibrator at the $1\%$ level.  However,
our preliminary results for Mars and Saturn suggest that with
additional analysis and observations, both of these sources may be
similarly useful in the future.
\section{Conclusions\label{sec:conc}} 
\wmap\ observes the planet Jupiter in two seasons a year, each of $\sim50$
days.  Ten seasons of Jupiter observations are used in this
paper to measure the in-flight beam patterns associated with each of
the multifrequency \wmap\ radiometers.  An accurate beam pattern
determination is critical for cosmological measurements.

Using the TOD, beam maps are formed from the Jupiter observations for
both the A-side and B-side optics.  The A-side fitting is improved
over previous analyses both by additional data and by extension of our
analysis techniques.  The B side is now directly fitted for the first
time.  The cutoff scale length of fitted primary mirror distortions is
reduced from previous analyses by a factor of $\sim 2$.  The
hybridization of beam models with beam data is optimized explicitly
with respect to error in the main beam solid angle.  We transform the
hybridized, symmetrized main beam profiles into harmonic space without
an intermediate spatial fitting function.

Although the beam transfer functions are statistically consistent with
earlier ones, a $\sim 1\%$ increase in solid angle is found for the V2
and W1--W4 DAs because of improved data and refinement of previous
analysis methods.  The uncertainty in the
beam transfer functions is decreased by a factor of $\sim 2$ relative to
previous \wmap\ beam analyses, demonstrating the success of continued
mission operations and continued progress from data analysis efforts. 
Extended operations and analysis will further reduce these
uncertainties.
\acknowledgements
We wish to thank Chris Barnes for his many contributions to the
study of \wmap\ beams, and especially for his development
of the physical optics model-fitting procedure.
We acknowledge
use of the HEALPix package \citep{gorski/etal:2005}.
The \wmap\ mission is made possible by the support of the Science Mission
Directorate Office at NASA Headquarters.  This research was additionally
supported by NASA grants NNG05GE76G, NNX07AL75G S01, LTSA03-000-0090,
ATPNNG04GK55G, and ADP03-0000-092.  This research has made use of
NASA's Astrophysics Data System Bibliographic Services.  
%
%

\clearpage
\begin{deluxetable}{cccc}
  \tablewidth{0pt}
  \tablecaption{Abbreviated Fit History for Side A \label{tab:hist-a}}
  \tablehead{
    \colhead{Primary} &
    \colhead{Primary} &
    \colhead{Secondary} &
    \colhead{} \\
    \colhead{$\kmax$} &
    \colhead{Modes} &
    \colhead{Modes} &
    \colhead{$\chi_\nu^2$}}
  \startdata
  12 & 122 & 30 & 1.15  \\
  14 & 158 & 30\tablenotemark{b} &  1.17 \\
  16 & 206 & 30\tablenotemark{b} &  1.11 \\
  18 & 262 & 30\tablenotemark{b} &  1.11 \\
  20 & 326 & 30\tablenotemark{b} &  1.12 \\
  22 & 386 & 30\tablenotemark{b} &  1.11 \\
  24 & 450 & 30\tablenotemark{b} &  1.05 \\
  \enddata
  \tablenotetext{a}{$\chi_\nu^2$ is approximate and indicates
  the progress of the fit.  Residuals are shown in
  Figure~\ref{fig:abresid}.}
  \tablenotetext{b}{Not refitted.}
\end{deluxetable}
\begin{deluxetable}{cccccccccl}
  \tablewidth{0pt}
  \tablecaption{Abbreviated Fit History for Side B \label{tab:hist-b}}
  \tablehead{
    \colhead{Primary} &
    \colhead{Primary} &
    \colhead{Secondary} &
    \colhead{} \\
    \colhead{$\kmax$} &
    \colhead{Modes} &
    \colhead{Modes} &
    \colhead{$\chi_\nu^2$}}
  \startdata
  4 & 22 & 0 & 2.50  \\
  6 & 38 & 9 & 1.26  \\
  6 & 38\tablenotemark{b} & 30 &  1.27 \\
  8 &  58 & 30\tablenotemark{b} & 1.22 \\
  10 &  90 & 30\tablenotemark{b} & 1.17 \\
  12 & 122 & 30\tablenotemark{b} & 1.17 \\
  14 & 158 & 30\tablenotemark{b} & 1.16 \\
  16 & 205 & 30\tablenotemark{b} & 1.15 \\
  18 & 262 & 30\tablenotemark{b} & 1.15 \\
  20 & 326 & 30\tablenotemark{b} & 1.15 \\
  22 & 386 & 30\tablenotemark{b} & 1.11 \\
  24 & 450 & 30\tablenotemark{b} & 1.07 \\
  \enddata
  \tablenotetext{a}{$\chi_\nu^2$ is approximate and indicates
  the progress of the fit.  Residuals are shown in
  Figure~\ref{fig:abresid}.}
  \tablenotetext{b}{Not refitted.}
\end{deluxetable}
\clearpage
\begin{deluxetable}{cccccc}
  \tablecaption{Main Beam Limits \label{tab:transrad}}
  \tablewidth{0pt}
  \tablehead{
    \colhead{} & 
    \colhead{3-Year Radius} & 
    \colhead{5-Year Radius} &
    \colhead{3-Year $B_\mathrm{thresh}$\tablenotemark{a}} & 
    \colhead{5-Year $B_\mathrm{thresh}$\tablenotemark{a}} &
    \colhead{$r_{50}$\tablenotemark{b}} \\
    \colhead{DA} & 
    \colhead{($\arcdeg$)} &
    \colhead{($\arcdeg$)} &
    \colhead{(dBi)} &
    \colhead{(dBi)} &
    \colhead{($\arcdeg$)}}
  \startdata
  K1 & 6.1  & 7.0 & 17 & 3 & 4.3\\
  Ka1& 4.6  & 5.5 & 17 & 4 & 3.5\\
  Q1 & 3.9  & 5.0 & 18 & 6 & 3.1\\
  Q2 & 3.9  & 5.0 & 18 & 6 & 3.1\\
  V1 & 2.5  & 4.0 & 19 & 8 & 2.5\\
  V2 & 2.5  & 4.0 & 19 & 8 & 2.4\\
  W1 & 1.7  & 3.5 & 20 & 11 & 1.8\\
  W2 & 1.7  & 3.5 & 20 & 11 & 1.7\\
  W3 & 1.7  & 3.5 & 20 & 11 & 1.7\\
  W4 & 1.7  & 3.5 & 20 & 11 & 1.8\\
  \enddata
  \tablenotetext{a}{Threshold in beam model gain relative to
    isotropic, below which model points are substituted for
    data points in two-dimensional hybrid beams.}
  \tablenotetext{b}{Radius in hybrid beam at which $50\%$ of
    radial profile points are from data and $50\%$ from beam model.}
\end{deluxetable}
\clearpage
\begin{deluxetable}{ccccc}
  \tabletypesize{\footnotesize}
  \tablewidth{0pt}
  \tablecolumns{5}
  \tablecaption{Main-Beam Solid Angles, Gains, 
    and $\Gamma$ for Combined Maps\label{tab:omega}}
  \tablehead{
    \colhead{} &
    \colhead{$\Omega^S$\tablenotemark{(a)}} & 
    \colhead{$\Delta(\Omega^S)/\Omega^S$\tablenotemark{(b)}} & 
    \colhead{$G_m$\tablenotemark{(c)}} & 
    \colhead{$\Gamma_\mathrm{ff}$\tablenotemark{(d)}}\\
    \colhead{DA} & 
    \colhead{(sr)} &
    \colhead{(\%)} &
    \colhead{(dBi)} &
    \colhead{($\mu$K Jy$^{-1}$)}}
  \startdata
  \cutinhead{For 10 Maps}
  K1  & $2.447\times10^{-4}$  & 0.7 & 46.97 & 262.2 \\
  Ka1 & $1.436\times10^{-4}$  & 0.5 & 49.41 & 211.8 \\
  Q1  & $8.840\times10^{-5}$  & 0.6 & 51.40 & 222.8 \\
  Q2  & $9.145\times10^{-5}$  & 0.6 & 51.29 & 216.6 \\
  V1  & $4.169\times10^{-5}$  & 0.4 & 54.85 & 214.1 \\
  V2  & $4.240\times10^{-5}$  & 0.4 & 54.75 & 205.7 \\
  W1  & $2.037\times10^{-5}$  & 0.4 & 57.97 & 184.7 \\
  W2  & $2.206\times10^{-5}$  & 0.4 & 57.67 & 168.7 \\
  W3  & $2.149\times10^{-5}$  & 0.5 & 57.77 & 176.6 \\
  W4  & $1.998\times10^{-5}$  & 0.5 & 57.95 & 186.7 \\
  \cutinhead{For 5 Maps}
  K   & $2.447\times10^{-04}$ & 0.7 & 46.97 & 262.7 \\
  Ka  & $1.436\times10^{-04}$ & 0.5 & 49.41 & 211.9 \\
  Q   & $8.993\times10^{-05}$ & 0.6 & 51.34 & 219.6 \\
  V   & $4.204\times10^{-05}$ & 0.4 & 54.80 & 210.1 \\
  W   & $2.098\times10^{-05}$ & 0.5 & 57.84 & 179.2 \\
  \enddata
  \tablenotetext{a}{Solid angle in azimuthally symmetrized beam.}
  \tablenotetext{b}{Relative error in $\Omega^S$.}
  \tablenotetext{c}{Forward gain $=$ maximum of gain relative to isotropic.}
  \tablenotetext{d}{Conversion factor to obtain flux density from
    \wmap\ antenna temperature, for a free-free spectrum.  The
    individual DA frequencies are taken from Table 3 of
    \citet{page/etal:2003e}.  The band average
    frequencies are taken to be 22.5, 32.7, 40.6, 60.7, and 93.05 GHz, for
    K--W respectively \citep{page/etal:2003}, and the band average $\Gamma_\mathrm{ff}$
    tabulated here are those used in the \wmap\ five-year source catalog
    \citep{wright/etal:prep}.}
\end{deluxetable}
\clearpage
\begin{deluxetable}{cccc}
  \tablewidth{0pt}
  \tablecaption{Five-Year Mean Jupiter Temperatures \label{tab:juptemp}}
  \tablehead{
    \colhead{DA} 
    & \colhead{\hspace{0.15in}$\nu_e^\mathrm{RJ}$\ \tablenotemark{a}\hspace{0.15in}} 
    & \colhead{\hspace{0.15in}$T$\tablenotemark{b}\hspace{0.15in}}
    & \colhead{\hspace{0.15in}$\sigma(T)$\tablenotemark{c}\hspace{0.15in}} \\
    & \colhead{\hspace{0.15in}(GHz)\hspace{0.15in}} 
    & \colhead{\hspace{0.15in}(K)\hspace{0.15in}} 
    & \colhead{\hspace{0.15in}(K)\hspace{0.15in}}}
  \startdata
  K1  & 22.8 & 135.2 & 0.93 \\
  Ka1 & 33.0 & 146.6 & 0.75 \\
  Q1  & 40.9 & 154.7 & 0.96 \\
  Q2  & 40.9 & 155.5 & 0.94 \\
  V1  & 61.0 & 165.0 & 0.80 \\
  V2  & 61.6 & 166.3 & 0.77 \\
  W1  & 93.8 & 172.3 & 0.78 \\
  W2  & 94.1 & 173.4 & 0.82 \\
  W3  & 93.2 & 174.4 & 0.87 \\
  W4  & 94.1 & 173.0 & 0.86 \\
  \enddata
  \tablenotetext{a}{Mean of A- and B-side values from Table 3 of
    \citet{page/etal:2003e}}
  \tablenotetext{b}{Brightness temperature
    calculated for a solid angle $\Omega_\mathrm{J}^\mathrm{ref}
    = 2.481 \times 10^{-8}$ sr at a fiducial distance $d_\mathrm{J}=5.2$ AU
    \citep{griffin/etal:1986}.  Temperature is with
    respect to blank sky; absolute brightness temperature is
    obtained by adding 2.2, 2.0, 1.9, 1.5, and 1.1 K in bands
    K, Ka, Q, V, and W, respectively \citep{page/etal:2003e}.}
  \tablenotetext{c}{Computed from errors
    in $\Omega^S$ (Table~\ref{tab:omega}) summed in quadrature with a
    calibration error of $0.2\%$.}
\end{deluxetable}
\clearpage
\begin{deluxetable}{cccccc}
  \tablecolumns{6}
  \tablewidth{0pt}
  \tablecaption{Jupiter Temperature Changes by Season \label{tab:juptime}}
  \tablehead{
    \colhead{Season\tablenotemark{a}} & \colhead{Start} & \colhead{End} 
    & \multicolumn{2}{c}{$\Delta T/T$ (\%)} & \colhead{$r/\overline{r}$\ \tablenotemark{d}} \\
    & & & \colhead{Mean\tablenotemark{b} } & \colhead{Scatter\tablenotemark{c}} & }
  \startdata
  1 & 2001/10/08 & 2001/11/22 &  0.16   & 0.12 & 0.96 \\
  3 & 2002/11/10 & 2002/12/24 & $-0.14$ & 0.23 & 0.97 \\
  4 & 2003/03/15 & 2003/04/29 & $-0.25$ & 0.49 & 0.99 \\
  5 & 2003/12/11 & 2004/01/23 & $-0.02$ & 0.24 & 1.00 \\
  6 & 2004/04/15 & 2004/05/30 & $-0.01$ & 0.30 & 1.01 \\
  7 & 2005/01/09 & 2005/02/21 & 0.03    & 0.19 & 1.02 \\
  8 & 2005/05/16 & 2005/07/01 & 0.01    & 0.27 & 1.02 \\
  9 & 2006/02/07 & 2006/03/24 & 0.05    & 0.19 & 1.02 \\
  10 & 2006/06/16 & 2006/08/02 & 0.08    & 0.35 & 1.02 \\
  \enddata
  \tablenotetext{a}{Season 2 omitted because Jupiter is in the Galactic
    plane.}
  \tablenotetext{b}{Mean of the percentage brightness temperature change among the
    DAs for each season, relative to the 5-year mean.}
  \tablenotetext{c}{$1\sigma$ scatter in the percentage temperature change among the
    DAs for each season.}
  \tablenotetext{d}{Mean Jupiter--\wmap\ distance for each season, relative to the 10-season
    mean$=$ 5.34 AU.}
\end{deluxetable}
\clearpage
\begin{deluxetable}{ccccc}
  \tabletypesize{\footnotesize}
  \tablewidth{0pt}
  \tablecolumns{3}
  \tablecaption{W-Band Observations of Mars and
    Saturn\label{tab:satmars}}
  \tablehead{
    \colhead{Julian Day\tablenotemark{a}} &
    \colhead{$T_\mathrm{planet}$\tablenotemark{b}} & 
    \colhead{Scatter\tablenotemark{c}} \\
    \colhead{$-2450000$} & 
    \colhead{(K)} &
    \colhead{(K)}}
  \startdata
  \cutinhead{Mars}
  2184  & \hspace{0.5in}  185 \hspace{0.5in} &   5.4 \\
  2775  & \hspace{0.5in}  199 \hspace{0.5in} &   6.2 \\
  2984  & \hspace{0.5in}  191 \hspace{0.5in} &   5.6 \\
  3583  & \hspace{0.5in}  208 \hspace{0.5in} &   4.4 \\
  3758  & \hspace{0.5in}  182 \hspace{0.5in} &   5.9 \\
  \cutinhead{Saturn}
  2178  & \hspace{0.5in} 161 \hspace{0.5in} &   2.2 \\
  2310  & \hspace{0.5in} 160 \hspace{0.5in} &   2.2 \\
  2562  & \hspace{0.5in} 162 \hspace{0.5in} &   2.2 \\
  2942  & \hspace{0.5in} 157 \hspace{0.5in} &   2.2 \\
  3321  & \hspace{0.5in} 150 \hspace{0.5in} &   2.2 \\
  3449  & \hspace{0.5in} 154 \hspace{0.5in} &   2.4 \\
  3700  & \hspace{0.5in} 143 \hspace{0.5in} &   2.2 \\
  3828  & \hspace{0.5in} 147 \hspace{0.5in} &   3.1 \\
  \enddata
  \tablenotetext{a}{Approximate mean time of observations in 
    each season.}
  \tablenotetext{b}{Mean of W band brightness temperatures from
    the \wmap\ DAs W1--W4, with respect to
    blank sky (see Table \ref{tab:juptemp}, note \emph{b}).  
    Fiducial solid angles
    are $7.156 \times 10^{-10}$ for Mars and 
    $5.101 \times 10^{-9}$ sr for Saturn 
    \citep{hildebrand/etal:1985}.}
  \tablenotetext{c}{$1\sigma$ scatter among the four W-band
    DAs.}
\end{deluxetable}
\clearpage
\clearpage
\clearpage\clearpage
\clearpage
\clearpage
\clearpage
\clearpage
\clearpage
\clearpage
\clearpage
\clearpage

\end{document}